\documentclass[twocolumn,showpacs,preprintnumbers,amsmath,amssymb]{revtex4}

\usepackage{graphicx}
\usepackage{dcolumn}
\usepackage{bm}

\begin{document}

\title{On Possible Implications of Self-Organization Processes
through \\ Transformation of Laws of Arithmetic into Laws of Space
and Time}

\author{Victor Korotkikh}
\email{v.korotkikh@cqu.edu.au}

\affiliation{School of Information and Communication Technology\\
CQUniversity, Mackay \\ Queensland 4740, Australia}

\begin{abstract}

In the paper we present results based on the description of complex
systems in terms of self-organization processes of prime integer
relations. Realized through the unity of two equivalent forms, i.e.,
arithmetical and geometrical, the description allows to transform
the laws of a complex system in terms of arithmetic into the laws of
the system in terms of space and time. Possible implications of the
results are discussed.

\end{abstract}


\maketitle

\section{Introduction}

In the paper we present results based on the description of complex
systems in terms of self-organization processes of prime integer
relations and discuss their possible implications.

Rather than space and time, the description suggests a new stage for
understanding and dealing with complex systems, i.e., the
hierarchical network of prime integer relations. It appears as the
structure built by the totality of the processes and existing
through the mutual consistency of its parts. Based on the integers
and controlled by arithmetic only the description can picture
complex systems by irreducible concepts alone and thus secure its
foundation. Remarkably, this raises the possibility to develop an
irreducible theory of complex systems
\cite{Korotkikh_1}-\cite{Korotkikh_9}.

In section II we give some basics of the description to present its
two equivalent forms and show that the description can work
arithmetically and geometrically all at once.

In the arithmetical form a complex system is characterized by
hierarchical correlation structures determined by self-organization
processes of prime integer relations. The correlation structures
operate through the relationships emerging in the formation of the
prime integer relations. Since a prime integer relation expresses a
law between the integers, the complex system is, in fact, governed
by the laws of arithmetic realized through the self-organization
processes of prime integer relations.

In the geometrical form the correlation structures of the complex
system are given by hierarchical structures of two-dimensional
geometrical patterns, as the processes become isomorphically
expressed in terms of their transformations. This geometrizes the
correlations as well as the laws of arithmetic the complex system is
determined by. As a result, the geometrization gives the body to the
correlations and to the laws of arithmetic to be characterized by
space and time as dynamics variables. This allows to transform the
laws of the complex system in terms of arithmetic into the laws of
the system in terms of space and time.

To have a picture of the hierarchical network in section III we
consider a process that can probe the hierarchical network on all
levels.

In section IV we discuss a scale-invariant property of the process
suggesting its effective representation, where the levels are
arranged into the groups of three successive levels with important
consequences. In particular, by using renormalizations in such a
group the process can be given by a series of approximations so that
the first term characterizes the process in a self-similar way to
the characterization at levels $1, 2$ and $3$.

Consequently, the process at these three levels provides a first
resolution picture of the hierarchical network, where the
correlation structure determined by the process is isomorphically
represented by a hierarchical structure of two-dimensional
geometrical patterns.

In section V we analyze the picture in more detail and represent the
hierarchical structure of geometrical patterns by using space and
time as dynamical variables. This allows to transform the laws of
the process in terms of arithmetic into the laws in terms of space
and time. As a result, local spacetimes of the elementary parts
become defined and we can consider how they appear to be related to
one another.

Remarkably, in the representation the elementary parts of the
correlation structure act as the carriers of the laws of arithmetic
with each elementary part carrying its own quantum of the laws. This
opens an important perspective to use elementary parts as quanta to
construct different laws and in section VI we consider how the laws
of arithmetic could be transformed into different forms by
constructing global spacetimes.

In section VII we discuss possible implications of the results.

\section{Basics of the Description}

The description of complex systems in terms of self-organization
processes of prime integer relations is realized through the unity
of two equivalent forms, i.e., arithmetical and geometrical
\cite{Korotkikh_1}-\cite{Korotkikh_9}.

In particular, in the geometrical form $N \geq 2$ elementary parts
$P_{10},...,P_{N0}$ as the initial building blocks in the formation
of a complex system are considered at level $0$. An elementary part
$P_{j0}, j = 1,...,N$ is given through a local reference frame
specified by two parameters $\delta_{j} > 0$ and $\varepsilon_{j} >
0$. The local reference frame is a setting to characterize the
elementary part $P_{j0}$ and accommodate the changes determined by
the formation. The reference frames are arranged to consider the
elementary parts $P_{10},...,P_{N0}$ simultaneously, yet each in its
own reference frame without information about the distances in space
and time.

The geometrical form requires the parameters to be the same
$$
\delta = \delta_{j}, \ \ \ \varepsilon = \varepsilon_{j}, \ \ \ j =
1,...,N,
$$
which are associated with dimensionless quantities of space and time
$$
\delta = \frac{\chi_{0}}{\chi_{min}}, \ \ \ \varepsilon =
\frac{\tau_{0}}{\tau_{min}},
$$
where $\chi_{0}$ and $\tau_{0}$ are the length scales of space and
time at level $0$, and $\chi_{min}$ and $\tau_{min}$ are
corresponding minimum length scales.

Since
$$
\chi_{0} \geq \chi_{min}, \ \ \ \tau_{0} \geq \tau_{min},
$$
we have
\begin{equation}
\label{SectionII.1} \delta \geq 1, \ \ \  \varepsilon \geq 1.
\end{equation}

In general, the parameters $\delta$ and $\varepsilon$ give us a
choice in setting the geometrical form and become the basic
constants, which can not be changed unless in all reference frames
of the elementary parts.

The state of an elementary part $P_{j0}, j = 1,...,N$ is determined
to characterize the geometry of a corresponding self-organization
process of prime integer relations at level $0$. In particular, the
state of the elementary part $P_{j0}$ can be given by the space
coordinate $X_{j0} = s_{j}\delta$, while the time coordinate
$T_{j0}$ changes independently by $\varepsilon$, where $s_{j} \in I$
and $I$ is a set of integers. The state of the elementary parts
$P_{10},...,P_{N0}$ can be specified by a sequence
$$
s = s_{1}...s_{N} \in I_{N},
$$
where $I_{N}$ is a set of sequences of length $N$, and represented
by a piecewise constant function.

In particular, let
$$
\rho_{m\varepsilon\delta}: s \rightarrow f
$$
be a mapping that associates a sequence $s = s_{1}...s_{N} \in
I_{N}$ with a function $f$, denoted $f =
\rho_{m\varepsilon\delta}(s)$, such that
$$
f(t) = s_{j}\delta, \ \ \ t \in [t_{m+j-1},t_{m+j}),\ \ \ j =
1,...,N,
$$
$$
f(t_{m+N}) = s_{N}\delta, \ \ \ t_{j} = j\varepsilon, \ \ \ j =
m,...,m + N
$$
and
$$
f^{[j]}(t_{m}) = 0, \ \ \ j = 1,2,... \ ,
$$
where $m$ is an integer.

Let
$$
I_{N}(Q_{1},...,Q_{k}) \subset I_{N}
$$
be the states of the elementary parts $P_{10},...,P_{N0}$ such that
if
$$
s = s_{1}...s_{N} \in I_{N}(Q_{1},...,Q_{k})
$$
and
$$
s' = s_{1}'...s_{N}' \in I_{N}(Q_{1},...,Q_{k}),
$$
then
$$
Q_{j} = f^{[j]}(t_{m+N}) = g^{[j]}(t_{m+N}), \ \ \ j = 1,...,k,
$$
but
$$
f^{[k+1]}(t_{m+N}) \neq g^{[k+1]}(t_{m+N}),
$$
where $f^{[j]}(t)$ and $g^{[j]}(t), \ t \in [t_{m},t_{m+N}], \ j =
1,...,k+1$ are the $j$th integrals of the functions $f =
\rho_{m\varepsilon\delta}(s)$ and $g =
\rho_{m\varepsilon\delta}(s')$ accordingly.

Importantly, the quantities $Q_{1},...,Q_{k}$ can define a complex
system and its formation from the elementary parts
$P_{10},...,P_{N0}$ \cite{Korotkikh_1}-\cite{Korotkikh_6}. In
particular, the states $I_{N}(Q_{1},...,Q_{k})$ of the elementary
parts $P_{10},...,P_{N0}$ can determine the states of the system
with the transitions preserving the quantities $Q_{1},...,Q_{k}$ and
thus the system itself. Since a system can be in one of the possible
states, it is not possible to predict which state will be actually
measured in any given case and thus the description provides the
statistical information about the system.

The integer code series \cite{Korotkikh_1}, as the origin of the
description, plays a crucial role. In fact, it is the key to prove
that in the transition between two states
$$
s = s_{1}...s_{N} \in I_{N}(Q_{1},...,Q_{k})
$$
and
$$
s' = s_{1}'...s_{N}' \in I_{N}(Q_{1},...,Q_{k})
$$
of the elementary parts $P_{10},...,P_{N0}$ the $Q_{1},...,Q_{k}$
quantities remain invariant
\begin{equation}
\label{SectionII.2} Q_{j} = f^{[j]}(t_{m+N}) = g^{[j]}(t_{m+N}), \ \
\ j = 1,..., k,
\end{equation}
but
$$
f^{[k+1]}(t_{m+N}) \neq g^{[k+1]}(t_{m+N}),
$$
if and only if the correlations between the parts in the formation
of the system are defined by $k$ Diophantine equations
$$
(m + N)^{k-1}\Delta s_{1} + ... + (m + 1)^{k-1}\Delta s_{N} = 0
$$
$$
. \qquad \qquad . \qquad  \qquad .  \qquad \qquad . \qquad \qquad .
$$
$$
(m + N)^{1}\Delta s_{1} + ... + (m + 1)^{1}\Delta s_{N} = 0
$$
\begin{equation}
\label{SectionII.3} (m + N)^{0}\Delta s_{1} + ... + (m +
1)^{0}\Delta s_{N} = 0
\end{equation}
and an inequality
\begin{equation}
\label{SectionII.4} (m + N)^{k}\Delta s_{1} + ... + (m +
1)^{k}\Delta s_{N} \neq 0,
\end{equation}
where $f = \rho_{m\varepsilon\delta}(s), \ g =
\rho_{m\varepsilon\delta}(s')$ and
$$
\Delta s_{1}...\Delta s_{N} = s_{1}' - s_{1},...,s_{N}' - s_{N}.
$$

Significantly, through the analysis of the Diophantine equations
$(\ref{SectionII.3})$ and inequality $(\ref{SectionII.4})$ certain
hierarchical structures can be revealed and interpreted as a result
of self-organization processes of prime integer relations (Figure
1). In fact, the processes give rise to the hierarchical structures
of prime integer relations, which entirely determine the correlation
structures (Figure 2) in control of the transition of the system
\cite{Korotkikh_1}-\cite{Korotkikh_6}.

The concept of prime integer relation captures that the prime
integer relations are built by the processes as indivisible wholes.
In particular, a prime integer relation is made by a process first
from integers and then prime integer relations from the levels
below. Remarkably, all the components of the prime integer relation
are necessary and sufficient for the prime integer relation to
exist. In other words, a prime integer relation can be seen as a
system itself, where each and every part is important for its
formation \cite{Korotkikh_1}-\cite{Korotkikh_6}.

In the description a correlation structure of a complex system is
determined by a self-organization process of prime integer
relations. In particular, under a self-organization process
elementary parts of level $0$ combine into parts of level $1$, which
in their turn compose more complex parts of level $2$ and so on. The
formation continues as long as the process can provide the
relationships for parts to be made. Notably, in the formation an
elementary part of a level transforms into an elementary part of a
larger part at the higher level.

\begin{figure}
\includegraphics[width=.52\textwidth]{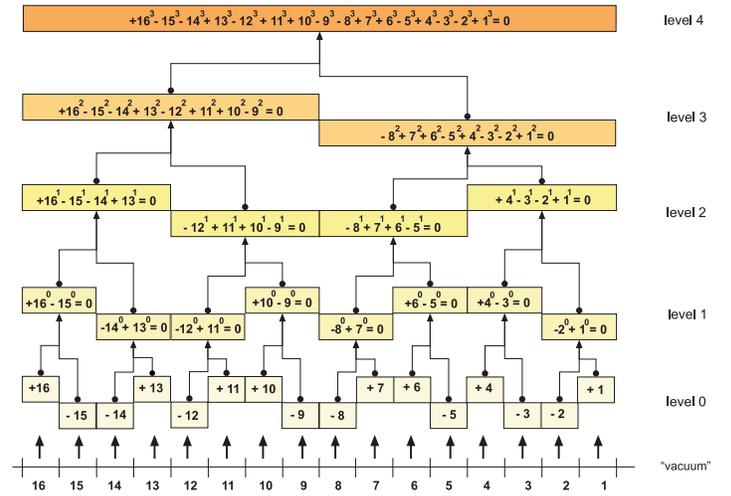}
\caption{\label{fig:two} The figure shows a hierarchical structure
of prime integer relations, which can be interpreted as a result of
a self-organization process. The process starts when integers $16,
13, 11, 10, 7, 6, 4, 1$ in the positive state and integers $15, 14,
12, 9, 8, 5, 3, 2$ in the negative state are generated from the
"vacuum" to appear at level $0$. The "vacuum" encodes all possible
processes to be realized in the construction of the hierarchical
network of prime integer relations. Since a prime integer relation
expresses a law between the integers, the structure can be seen as a
hierarchical set of laws of arithmetic.}
\end{figure}

Remarkably, the correlation structures of the transitions entirely
define the system. Indeed, they strictly determine the changes
between the states so that the quantities and thus the system remain
the same. As each of the correlation structures is ready to exercise
its own scenario and there is no mechanism specifying which of them
is going to take place, an intrinsic uncertainty about the system
exists. At the same time the information about the correlation
structures can be used to evaluate the probability of a state
observable to take each of the possible outcomes.

Importantly, the Diophantine equations $(\ref{SectionII.3})$ have no
reference to the distances between the parts in space and time, and
thus suggest that the correlations are nonlocal in character.
Therefore, according to the description, parts of a complex system
may be far apart in space and time and yet remain interconnected
with instantaneous effect on each other through the prime integer
relations. In fact, through the prime integer relations the
elementary parts receive information instantaneously, wherever they
happen to be in the correlation structure.

Moreover, since a prime integer relation expresses a law between the
integers, consequently, a complex system become governed by the laws
of arithmetic realized through the self-organization processes of
prime integer relations.

In the geometrical form the correlation structures of a complex
system become isomorphically represented by hierarchical structures
of two-dimensional patterns \cite{Korotkikh_1}-\cite{Korotkikh_3}.
Importantly, this geometrizes the correlations as well as the laws
of arithmetic the complex system is determined by and allows to
consider their representations in terms of space and time. As a
result, the space and time appear as a manifestation of the prime
integer relations and thus the integers
\cite{Korotkikh_6}-\cite{Korotkikh_9}.

Significantly, the geometrization allows to transform the laws of a
complex system in terms of arithmetic into the laws of the system in
terms of space and time. In particular, this specifies the state of
an elementary part at level $0$.

In the arithmetical form a self-organization process starts when
integers are generated from the "vacuum" to appear at level $0$ with
an integer in the positive or negative state. As a result, an
elementary part, depending on the state of a corresponding integer,
becomes positive or negative. At level $0$, where there are no
relationships between the integers to provide interactions for the
elementary parts, an elementary part is not forced to move and can
be defined in the state of rest. In the geometrical form the
boundary curve of the two-dimensional pattern of an elementary part
at level $0$ is specified by a constant function. Therefore, the
boundary curve of the geometrical pattern can be represented by the
space coordinate of the elementary part as a constant, given by the
value of the function, while the time coordinate can change
independently by $\varepsilon$.

Remarkably, the state of an elementary part at level $0$ reveals a
parallel with the law of inertia postulated in classical mechanics
\cite{Galilei_1}-\cite{Mach_1}. In our description the state of an
elementary part is defined by the laws of arithmetic realized
through the processes at level $0$.

Rather than space and time, the description suggests a new stage for
understanding and dealing with complex systems, i.e., the
hierarchical network of prime integer relations. It appears as the
structure built by the totality of the processes and existing
through the mutual consistency of its parts
\cite{Korotkikh_1}-\cite{Korotkikh_6}.

In the hierarchical network an event takes place through a
correlation structure as a basic causal element, which defines the
notion of simultaneity in the description. In particular, once a
correlation structure of a complex system turns to be operational,
then through the prime integer relations the parts, irrespective of
the distances and levels, become all instantaneously connected, and
the correlations are simultaneously realized for the geometrical
patterns to be in full agreement with the prime integer relations.

Notably, the effect of the correlations on an elementary part is
given by its geometrical pattern and can be different for elementary
parts. However, for each elementary part it is exactly as required
to preserve the system. In particular, in the representation of the
correlations with space and time as dynamic variables it may turn
out that the clocks of elementary parts tick differently, yet
exactly as determined by the prime integer relations
\cite{Korotkikh_6}-\cite{Korotkikh_9}.

Strictly controlled by arithmetic the prime integer relations as
well as their geometrical patterns can not be changed even a bit and
thus stand as rigid bodies interconnected by the processes. As a
result, the hierarchical network can be used as an absolute
reference for motion. Namely, in the hierarchical network the
position and motion of a complex system can be specified by the
processes in control of the correlation structures. Moreover, in
space and time the motion of an elementary part relative to this
absolute frame of reference can be defined by the representation of
the correlations through the geometrical pattern.

\begin{figure}
\includegraphics[width=.50\textwidth]{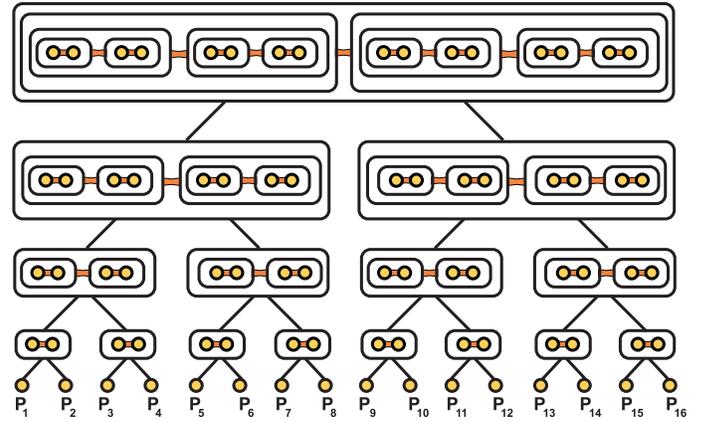}
\caption{\label{fig:two} Correlation structure corresponding to the
hierarchical structure of prime integer relations. The elementary
parts at level $0$ shown as $P_{1},...,P_{16}$ are the initial
building blocks in the formation of the hierarchical correlation
structure. The structure is built across different levels as under
the self-organization process the elementary parts combine into
parts, which in their turn compose more complex parts and so on. The
formation continues as long as the prime integer relations provide
the relationships.}
\end{figure}

Importantly, based on the integers and controlled by arithmetic only
the description can picture complex systems by irreducible concepts
alone and thus secure its foundation. This raises the possibility to
develop an irreducible theory of complex systems
\cite{Korotkikh_1}-\cite{Korotkikh_9}.

In the next section we consider a process that can probe the
hierarchical network on all levels and thus may provide information
about it as a whole.

\section{Making a Picture of the Hierarchical Network}

Let us consider a self-organization process of prime integer
relations that can determine a correlation structure of a complex
system with the changes
$$
\Delta s_{1}...\Delta s_{N} = s_{1}' - s_{1},...,s_{N}' - s_{N}
$$
between two states
$$
s = s_{1}...s_{N}, \ \ \ s' = s_{1}'...s_{N}'
$$
of the elementary parts $P_{10},...,P_{N0}$ specified by the PTM
(Prouhet-Thue-Morse) sequence
$$
\eta = +1-1-1+1-1+1+1-1 ... = \eta_{1}...\eta_{N} ... \ ,
$$
where $\Delta s_{j} = \eta_{j}, \ j = 1,...,N$.

Significantly, as the number $N$ of the elementary parts
$P_{10},...,P_{N0}$ increases, the process can probe the
hierarchical network on all its levels and, in particular, reach
level $log_{2}N$, when $N = 2^{k}, k = 1,2,... \ $
\cite{Korotkikh_1}. In this case the Diophantine equations
$(\ref{SectionII.3})$ and inequality $(\ref{SectionII.4})$ become
respectively
$$
\eta_{1}N^{k-1} + ... + \eta_{N-1}2^{k-1} + \eta_{N}1^{k-1} = 0
$$
$$
. \qquad \qquad . \qquad  \qquad .  \qquad \qquad .
$$
$$
\eta_{1}N^{1} + ... + \eta_{N-1}2^{1} + \eta_{N}1^{1} = 0
$$
\begin{equation}
\label{SectionIII.1} \eta_{1}N^{0} + ... + \eta_{N-1}2^{0} +
\eta_{N}1^{0} = 0
\end{equation}
and
\begin{equation}
\label{SectionIII.2} \eta_{1}N^{k} + ... + \eta_{N-1}2^{k} +
\eta_{N}1^{k} \neq 0,
\end{equation}
where $m = 0$. For example, when $N = 16$ we can explicitly write
down $(\ref{SectionIII.1})$ and $(\ref{SectionIII.2})$ as
$$
+16^{3} - 15^{3} - 14^{3} + 13^{3} - 12^{3} + 11^{3} + 10^{3} -
9^{3}
$$
$$
-8^{3} + 7^{3} + 6^{3} - 5^{3} + 4^{3} - 3^{3} - 2^{3} + 1^{3} = 0
$$
$$
+16^{2} - 15^{2} - 14^{2} + 13^{2} - 12^{2} + 11^{2} + 10^{2} -
9^{2}
$$
$$
-8^{2} + 7^{2} + 6^{2} - 5^{2} + 4^{2} - 3^{2} - 2^{2} + 1^{2} = 0
$$
$$
+16^{1} - 15^{1} - 14^{1} + 13^{1} - 12^{1} + 11^{1} + 10^{1} -
9^{1}
$$
$$
-8^{1} + 7^{1} + 6^{1} - 5^{1} + 4^{1} - 3^{1} - 2^{1} + 1^{1} = 0
$$
$$
+16^{0} - 15^{0} - 14^{0} + 13^{0} - 12^{0} + 11^{0} + 10^{0} -
9^{0}
$$
\begin{equation}
\label{SectionIII.Iden}
-8^{0} + 7^{0} + 6^{0} - 5^{0} + 4^{0} -
3^{0} - 2^{0} + 1^{0} = 0
\end{equation}
and
$$
+16^{4} - 15^{4} - 14^{4} + 13^{4} - 12^{4} + 11^{4} + 10^{4} -
9^{4}
$$
$$
-8^{4} + 7^{4} + 6^{4} - 5^{4} + 4^{4} - 3^{4} - 2^{4} + 1^{4} \neq
0.
$$

Importantly, the resulting system of integer identities
(\ref{SectionIII.Iden}) can be seen as a hierarchical set of laws of
arithmetic (Figure 1).

The self-organization process starts as integers $N,...,1$ make a
transition from the "vacuum" to level $0$, where each integer
acquires the state, positive or negative, depending on the sign of
the corresponding element in the PTM sequence. Then the integers
combine into pairs and make up the integer relations of level $1$.
Following a single organizing principle
\cite{Korotkikh_1}-\cite{Korotkikh_3} the process continues as long
as arithmetic allows the integer relations of a level to form the
integer relations of the higher level. Notably, in each of the
integer relations all components it is made of are necessary and
sufficient and this is the reason why we call the integer relations
prime.

As long as the integers of level $l = 0$ or the prime integer
relations of level $l = 1,...,k-1$ can form the prime integer
relations of level $l+1$, the relationships for the parts of level
$l$ to compose the parts of level $l+1$ are provided (Figure 2). As
a result, the elementary parts $P_{1l},...,P_{Nl}$ of level $l$
transform to become the elementary parts $P_{1,l+1},...,P_{N,l+1}$
of more complex parts of level $l+1$. Within a part the elementary
parts can effect each other through the relationships provided by
the prime integer relation thus making the relationships
instrumental in the preservation of the part.

Remarkably, in the geometrical form the self-organization process
become isomorphically represented by transformations of
two-dimensional patterns (Figure 3). In particular, under the
isomorphism a prime integer relation turns into the geometrical
pattern, which can be viewed as the prime integer relation itself,
but only expressed geometrically.

At level $0$ the geometrical patterns corresponding to the integers
and the elementary parts $P_{10},...,P_{N0}$ are specified by a
piecewise constant function
$$
\Psi^{[0]}_{1} = \Psi_{1} = \rho_{0\varepsilon\delta}(\eta), \ \eta
\in I_{N}
$$
such that
$$
\Psi^{[0]}_{1}(t) = \eta_{j}\delta, \ \ \ t_{j-1} \leq t < t_{j},
$$
$$
t_{j} = j\varepsilon, \ \ \ j = 1,...,N.
$$

In our description the geometrical pattern of an elementary part
$P_{j0}, j = 1,...,N$ is defined by the region enclosed by its
boundary curve, i.e., the graph of the function $\Psi_{1}^{[0]}(t),
t_{j-1} \leq t < t_{j}$, the vertical lines $t = t_{j-1}, t = t_{j}$
and the $t$-axis. In their turn the space and time coordinates of
the elementary part $P_{j0}$ are defined by the representation of
the boundary curve of the geometrical pattern.

In particular, in the transition from one state into another at the
moment $T_{j0}(t_{j-1}) = 0$ of the local time the space coordinate
$X_{j0}(t_{j-1})$ of the elementary part $P_{j0}$ changes by
$$
\Delta X_{j0} = \Psi^{[0]}_{1}(t_{j-1}) = \eta_{j}\delta
$$
and then stay as it is, while the time coordinate $T_{j0}(t)$
changes independently with the length of the boundary curve by
$\Delta T_{j0} = \varepsilon$.

At level $l = 1,...,k$ the geometrical pattern of the $j$th part $j
= 1,...,2^{k-l}$ is defined by the region enclosed by the boundary
curve, i.e., the graph of the function
$$
\Psi^{[l]}_{1}(t), \ \ \ 2^{l}(j-1)\varepsilon \leq t \leq
2^{l}j\varepsilon,
$$
and the $t$-axis.

Due to the isomorphism, as the integers at level $l = 0$ or the
prime integer relations at level $l = 1,...,k-1$ form the prime
integer relations at level $l+1$, under the integration of the
function $\Psi^{[l]}_{1}(t), 0 \leq t \leq t_{N}$ the geometrical
patterns of the parts at level $l$ transform into the geometrical
patterns of the parts at level $l+1$ (Figure 3). As a result, the
geometrical pattern of an elementary part $P_{jl}, j = 1,...,N$ at
level $l$, i.e., the region enclosed by the graph of the function
$\Psi_{1}^{[l]}(t), t_{j-1} \leq t \leq t_{j}$, the vertical lines
$t = t_{j-1}, t = t_{j}$ and the $t$-axis, transforms into the
geometrical pattern of the elementary part $P_{j,l+1}$ at level
$l+1$, i.e., the region enclosed by the graph of the function
$\Psi_{1}^{[l+1]}(t), t_{j-1} \leq t \leq t_{j}$, the vertical lines
$t = t_{j-1}, t = t_{j}$ and the $t$-axis.

Importantly, the $l$th integral $\Psi^{[l]}_{1},\ l = 1,...,k$
inherits the signature of the PTM sequence \cite{Korotkikh_3} in the
sense that at level $l$
\begin{equation}
\label{SectionIII.3} \Psi^{[l]}_{1}(2^{l}(j-1)\varepsilon + t) =
\eta_{j}\Psi^{[l]}_{1}(t),
\end{equation}
where $0 \leq t \leq 2^{l}\varepsilon, \ j = 1,..., 2^{k-l}$ (Figure 3).

\begin{figure}
\includegraphics[width=.49\textwidth]{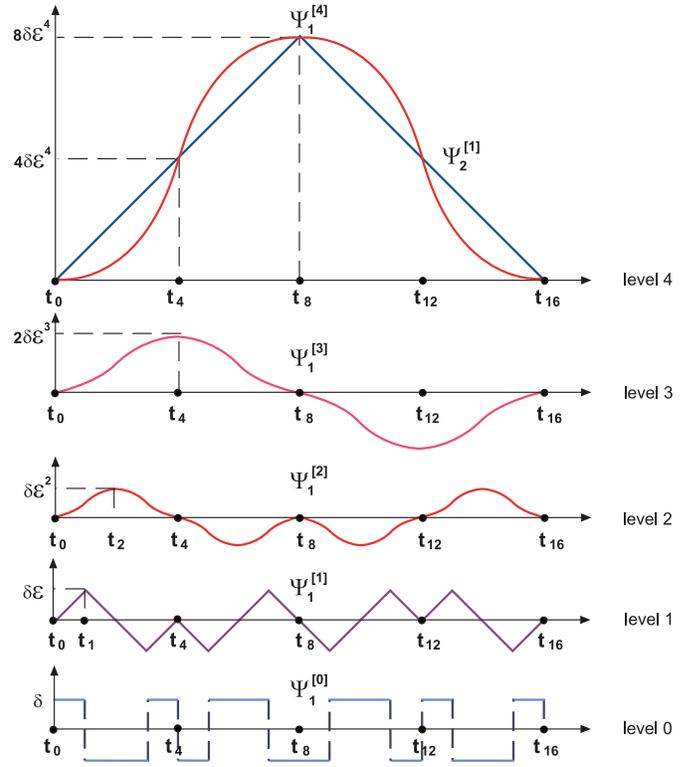}
\caption{\label{fig:one} Hierarchical structure of geometrical
patterns corresponding to the process. The isomorphism allows us to
see a direct picture of the geometrical patterns, prime integer
relations and correlations all together. For example, through the
integration of the function $\Psi_{1}^{[l]}, l = 1,2,3$ we can see
how the prime integer relations of level $l$ become the prime
integer relations of the higher level. The functions
$\Psi_{1}^{[1]},...,\Psi_{1}^{[4]}$ express the prime integer
relations and can be used to measure the effect of correlations on
the elementary parts. Remarkably, through the prime integer
relations arithmetic defines how the boundary curves of the
geometrical patterns must be curved. Notably, all geometrical
patterns are symmetrical. Moreover, the symmetries are
interconnected through the process and belong to one large symmetry
of the hierarchical network.}
\end{figure}

Now, let us consider how the geometrical pattern can be
characterized. In this regard, two characteristics of the
geometrical pattern of the $j$th part $j = 1,...,2^{k-l}$ at level
$l = 1,...,k$ are especially important.

First, it is the base $D_{jl}$, i.e., the length of the line segment
with the endpoints at $(2^{l}(j-1)\varepsilon,0)$ and
$(2^{l}j\varepsilon,0)$, and thus
$$
D_{jl} = 2^{l}\varepsilon.
$$

Second, it is the height $H_{jl}$, i.e., the length of the line
segment with one endpoint at the center
$(2^{\frac{3}{2}l}j\varepsilon,0)$ of the geometrical pattern and
the other at the point $(2^{\frac{3}{2}l}j\varepsilon,
\Psi^{[l]}_{1}(2^{\frac{3}{2}l}j\varepsilon))$. The height $H_{jl}$
is equal to the extremum of the function $\Psi^{[l]}_{1}(t)$ on the
interval
$$
2^{l}(j-1)\varepsilon \leq t \leq 2^{l}j\varepsilon,
$$
which is attained at the center of the geometrical pattern and hence
$$
H_{jl} = \vert \Psi^{[l]}_{1}(2^{\frac{3}{2}l}j\varepsilon)\vert.
$$

It turns out that \cite{Korotkikh_3}
$$
D_{l} = D_{jl}, \ \ \ H_{l} = H_{jl},
$$
$$
j = 1,...,2^{k-l}, \ \ \ l = 1,...,k.
$$

Significantly, by using the base $D_{jl}$ and the height $H_{jl}$
alone we can find the area of the geometrical pattern. In fact,
although the geometrical pattern of the $j$th part $j =
1,...,2^{k-l}$ at level $l = 2,...,k$ is not a triangle, yet, due to
the condition (\ref{SectionIII.3}), the boundary curve, as the graph
of the function $\Psi^{[l]}_{1}$, provides a remarkable property:
the area $A_{jl}$ of the geometrical pattern can be simply
calculated by
\begin{equation}
\label{SectionIII.4}  A_{l} = A_{jl} = \frac{D_{l}H_{l}}{2} =
\frac{D_{jl}H_{jl}}{2}.
\end{equation}

As under the process the geometrical patterns of two parts at level
$l = 0,...,k-1$ transform into the geometrical pattern of the part
at level $l+1$, the transformation connects the geometrical patterns
and thus their characteristics with consequences for the parts at
two different levels.

In particular, the base $D_{l}$ of the geometrical pattern of a part
at level $l = 1,...,k$ equals the sum
$$
D_{l} = D_{l-1,left} + D_{l-1,right} = 2D_{l-1}
$$
of the bases of two geometrical patterns of the parts, while each
geometrical pattern at level $l-1$ has the base
$$
D_{l-1} = 2^{l-1}\varepsilon.
$$

By using the fundamental theorem of calculus we can find that the
areas $A_{l-1}$ of the geometrical patterns of two parts at level
$l-1$ equals the height $H_{l}$ of the geometrical pattern of the
part they produce at level $l = 1,...,k$
\begin{equation}
\label{SectionIII.5} H_{l} = A_{l-1}.
\end{equation}

From (\ref{SectionIII.4}) and (\ref{SectionIII.5}) we can obtain a
recursive formula
$$
H_{l} = A_{l-1} = \frac{D_{l-1}H_{l-1}}{2}
$$
connecting the heights of the geometrical patterns of levels $l =
1,...,k$ and $l-1$ and use it to express the area of the geometrical
pattern of a part at level $l = 0,...,k$ as
$$
A_{l} = 2^{\frac{l(l-1)}{2}}\varepsilon^{l+1}\delta.
$$

Moreover, we can find the difference between the area of the
geometrical pattern of a part at level $l = 1,...,k$ and the sum of
the areas of the geometrical patterns of the parts at level $l-1$
from which the part is made of
$$
\Delta A_{l,l-1} = A_{l} - 2A_{l-1}
$$
$$
= 2^{\frac{l(l-1)}{2}}\varepsilon^{l+1}\delta - 2 \cdot
2^{\frac{(l-1)(l-2)}{2}}\varepsilon^{l}\delta
$$
$$
= 2^{\frac{l(l-3)}{2}}\varepsilon^{l}\delta(2^{l}\varepsilon -
2^{2}).
$$

Since, according to (\ref{SectionII.1}), $\varepsilon \geq 1$, then
the area of the geometrical pattern for all levels when $l \geq 2$
is greater than the sum of the areas of the geometrical patterns it
is composed of
\begin{equation}
\label{SectionIII.6} \Delta A_{l,l-1} =
2^{\frac{l(l-3)}{2}}\varepsilon^{l}\delta(2^{l}\varepsilon - 2^{2})
> 0,
\end{equation}
except when $l = 2, \ \varepsilon = 1$, we get
$$
\Delta A_{21} = 2^{-1}\delta(2^{2} - 2^{2}) = 0.
$$

Notably, in the formation from level $0$ to level $1$ the difference
between the areas is
\begin{equation}
\label{SectionIII.7} \Delta A_{10} = \varepsilon^{2}\delta -
2\varepsilon\delta = \varepsilon\delta(\varepsilon - 2)
\end{equation}
and, therefore, $\Delta A_{10} > 0$ when $\varepsilon > 2$, $\Delta
A_{10} = 0$ when $\varepsilon = 2$, but $\Delta A_{10} < 0$ when
$\varepsilon < 2$.

Consequently, for $l = 1, \ \varepsilon \geq 2$ and $l \geq 2$ when
two geometrical patterns combine, the area of the geometrical
pattern they produce can only stay the same or increase, but can not
decrease. However, for $l = 1, \ \varepsilon < 2$ the area indeed
decreases.

Remarkably, a prime integer relation can be seen as a
multifunctional entity. Two functions of the prime integer relation
are especially important. They combine the characterization of an
elementary part in the hierarchical network in terms of information
and the characterization of the elementary part in space and time in
terms of energy.

First, a prime integer relation can be seen as a storage as well as
a carrier of information. Namely, a prime integer relation does
contain information and in the realization of the correlations
communicates the information to the elementary parts. Importantly,
this function of the prime integer relation can be entirely
expressed by the geometrical pattern, where the area of the
geometrical pattern can be associated with the amount of information
transmitted to the elementary parts and the area of the geometrical
pattern of an elementary part can be associated with the amount of
information received by the elementary part.

As a result, a part at level $l = 1,...,k$ can be characterized by
the information or the entropy $S_{l}$ of a corresponding prime
integer relation and measured by the area $A_{l}$ of the geometrical
pattern. Therefore, for the entropy $S_{l}$ of a part at level $l =
1,...,k$ we obtain
\begin{equation}
\label{SectionIII.8} S_{l} = A_{l}.
\end{equation}

Second, a prime integer relation can be also seen as a source of
energy. Namely, in the hierarchical network the law governing a part
is actually the law of arithmetic given by a corresponding prime
integer relation. Significantly, in the description the law of
arithmetic can be fully expressed by the geometrical pattern and
written in terms of the variables of its quantitative
representation.

In particular, once the space and time coordinates of the elementary
part are defined to encode the boundary curve and the area of the
geometrical pattern is associated with the energy of the elementary
part, the representation of the geometrical pattern become
completed. Therefore, in the space and time representation the
energy $E_{l}$ of a part at level $l = 1,...,k$ can be defined by
the sum of the energies of the elementary parts and thus to be equal
to the area of the geometrical pattern
\begin{equation}
\label{SectionIII.8A} E_{l} = A_{l}.
\end{equation}

Since through the geometrical pattern the motion of the elementary
parts is fully determined by the prime integer relation, the prime
integer relation can be seen as a source of energy making the motion
of the elementary parts possible.

From (\ref{SectionIII.8}) and (\ref{SectionIII.8A}), we obtain
$$
S_{l} = E_{l}
$$
and thus that the entropy of the part equals its energy. However, it
should be mentioned that these two quantities characterize the part
in two different arenas, i.e., the hierarchical network and space
and time accordingly.

Notably, the condition (\ref{SectionIII.8}) shows that the entropy
of a prime integer relation is proportional to the surface area of
the geometrical pattern. Therefore, the condition
(\ref{SectionIII.8}) reproduces the well-known connection between
the entropy of a black hole and the area of its surface
\cite{Bekenstein_1},\cite{Hawking_1}, but in its own terms.

Now, let us express the difference between the entropy of a part at
level $l = 1,...,k$ and the sum of the entropies of the parts at
level $l-1$ the part is made of
$$
\Delta S_{l,l-1} = S_{l} - 2S_{l-1} = A_{l} - 2A_{l-1}
$$
$$
= \Delta A_{l,l-1} =
2^{\frac{l(l-3)}{2}}\varepsilon^{l}\delta(2^{l}\varepsilon - 2^{2}).
$$

As a result, from (\ref{SectionIII.6}) and (\ref{SectionIII.7}) we
obtain that for $l = 1, \ \varepsilon \geq 2$ and $l \geq 2$, when
two parts combine, the entropy of the part they compose can only
stay the same or increase, but can not decrease. More significantly,
however, according to (\ref{SectionIII.7}), we find that for $l = 1,
\ \varepsilon < 2$ the entropy, in fact, decreases.

Therefore, the description might open a new way to explain the
second law of thermodynamics \cite{Carno_1}-\cite{Boltzmann_1}.
Moreover, by revealing the special case when the entropy decreases,
the description raises the possibility that the second law of
thermodynamics can loose its generality and appear as a
manifestation of a more fundamental entity, i.e., the
self-organization processes of prime integer relations and thus
arithmetic.

Likewise, for the energy we get
$$
\Delta E_{l,l-1} = E_{l} - 2E_{l-1} = A_{l} - 2A_{l-1}
$$
\begin{equation}
\label{SectionIII.8B} = \Delta A_{l,l-1} =
2^{\frac{l(l-3)}{2}}\varepsilon^{l}\delta(2^{l}\varepsilon - 2^{2}).
\end{equation}

Importantly, this means that under the process, except for $l = 1, \
\varepsilon \leq 2$ and $l = 2, \ \varepsilon = 1$, the energy of a
part is greater than the sum of the energies of the parts it is made
of and thus with each next level the energy increases. However, the
most striking finding from (\ref{SectionIII.8B}) is that the energy
can be simply lost, when $l = 1, \ \varepsilon < 2$.

The following extremum principle can be formulated: for given
$\varepsilon$ and $\delta$ in the formation of a part of a level
from the parts of the lower level the energy of the part has to be
extremized under the constraint of the prime integer relation.

Therefore, in the description arithmetic could be associated with a
source of energy controlled through the processes with the
consequences determined by the geometrical form. This source of
energy may be already observed through dark energy and matter
\cite{Riess_1},\cite{Perlmutter_1}, yet, it would be a completely
different story to be able to use it for technological advances.

Moreover, arithmetic through a prime integer relation determines the
distribution of the energy between the elementary parts and,
consequently, produces a discrete energy spectrum. In this regard,
it is important to note that a prime integer relation and thus its
geometrical pattern are very sensitive and can not be changed.
Indeed, even a minor change to the geometrical pattern results in
the breaking of the relationships provided by the prime integer
relation and thus the part falls apart. Therefore, the areas of the
geometrical patterns of the elementary parts and their energies have
to be absolutely fixed \cite{Korotkikh_6}-\cite{Korotkikh_9}.

For example, it can be found that the energy spectrum of the
elementary parts $P_{14},...,P_{16,4}$ at level $4$ is a set of
quantized values
$$
E(P_{14}),..., E(P_{84}) =
$$
$$
\frac{1}{120}, \frac{29}{120},
\frac{149}{120}, \frac{361}{120}, \frac{599}{120}, \frac{811}{120},
\frac{931}{120}, \frac{959}{120},
$$
where
$$
E(P_{j4}) = E(P_{17-j,4}), \ \ \ j = 1,...,8
$$
and $\varepsilon = 1, \ \delta = 1$. Clearly, the energy $E(P_{j4})$
of an elementary part $P_{j4}, j = 1,...,16$ can be given by the
equation
\begin{equation}
\label{SectionIII.9} E(P_{j4}) = h_{4}\nu(P_{j4}),
\end{equation}
where $h_{4} = \frac{1}{120}$ and $\nu(P_{j4})$ is an integer.

Similarly, the energy spectrum of the elementary parts
$P_{15},...,P_{32,5}$ at level $5$ is a set of quantized values
$$
E(P_{15}),..., E(P_{16,5}) =
$$
$$
\frac{1}{720}, \frac{61}{720}, \frac{539}{120}, \frac{2039}{720},
\frac{4919}{720}, \frac{9179}{720}, \frac{14461}{720},
\frac{20161}{720},
$$
$$
\frac{25919}{720}, \frac{31619}{720}, \frac{36901}{120},
\frac{41161}{720},
$$
$$
\frac{44041}{720}, \frac{45541}{720},
\frac{46019}{720}, \frac{46079}{720},
$$
where
$$
E(P_{j5}) = E(P_{33-j,5}), \ \ \ j = 1,...,16
$$
and $\varepsilon = 1, \ \delta = 1$. In this case the energy
$E(P_{j5})$ of an elementary part $P_{j5}, j = 1,...,32$ can given
by the equation
\begin{equation}
\label{SectionIII.10} E(P_{j5}) = h_{5}\nu(P_{j5}),
\end{equation}
where $h_{5} = \frac{1}{720}$ and $\nu(P_{j5})$ is an integer.

Interestingly, many of the integers in (\ref{SectionIII.9}) and
(\ref{SectionIII.10}) are actually prime numbers. It seems like to
make the elementary parts different, arithmetic tries to define
their energies by using prime numbers. Note that
(\ref{SectionIII.9}) and (\ref{SectionIII.10}) look similar to the
Planck's equation \cite{Planck_1}.

The two-dimensional character of the geometrical pattern suggests
another important way to express the energy $E(P_{jl})$ of an
elementary part $P_{jl}, j = 1,...,N, \ l = 0,...,k$. In particular,
let $C^{2}_{l}$ be a unit of two-dimensional area at level $l$,
then, since the energy $E(P_{jl})$ of an elementary part $P_{jl}$ is
given by the area of the geometrical pattern, we can write an
equation
\begin{equation}
\label{SectionIII.11}
E(P_{jl}) = M(P_{jl})C^{2}_{l},
\end{equation}
which introduces the mass $M(P_{jl})$ of the elementary part. In
symbolic appearance the equation looks similar to the Einstein's
formula \cite{Einstein_1}. Importantly, by using the equation
(\ref{SectionIII.11}) we can explain why the mass of the elementary
part has the value that it does and not even slightly otherwise.

\section{From a Scale Invariance
to a Picture of the Hierarchical Network}

In the hierarchical network of prime integer relations the process
can progress through all levels and thus may be used to provide
information about it as a whole. In fact, the following effective
representation of the process allows us to obtain a first resolution
picture of the hierarchical network.

The representation is based on a scale-invariant property of the
process suggesting to arrange the levels into the groups of three
successive levels. In particular, by using renormalizations in such
a group the process can be given by a series of approximations,
where the first term of the series characterizes the process in a
self-similar way to the characterization at levels $1, 2$ and $3$,
and each next term reveals the process at a finer resolution. In
other words, while the higher the level the process reaches to, the
more complex it becomes with more terms in the series, yet the first
term is always characterizes the process self-similarly to its
characterization at levels $1, 2$ and $3$.

More specifically, in the representation the levels are considered
through the groups of three consecutive levels
$$
\Pi_{1},...,\Pi_{p}, \ \ \ p = 1,2,... \ , \ \ \ N = 2^{3p+1}
$$
and in a group $\Pi_{l}, l = 1,...,p$ of levels
$$
3(l - 1) + 1, \ 3(l - 1) + 2, \ 3(l - 1) + 3
$$
the process can be specified by a series of functions
$$
\Psi^{[1]}_{l}, \Psi^{[4]}_{l-1},..., \Psi^{[3(l-1)+1]}_{1},
$$
where the function $\Psi^{[3(l - j) + 1]}_{j}, j = 1,...,l$ encodes
the formation of the parts at level $3(l - 1) + 1$ from the parts as
basic elements at level $3(j - 1) + 1$. The representation
demonstrates two important properties that are based on the
condition (\ref{SectionIII.3}).

First, the characterization of the process by using functions
$$
\Psi^{[1]}_{l}, \Psi^{[2]}_{l}, \Psi^{[3]}_{l}
$$ 
in a group $\Pi_{l}, l = 2,...,p$ of levels
$$
3(l - 1) + 1, \ 3(l - 1) + 2, \ 3(l - 1) + 3
$$
turns out to be self-similar to the characterization of the process
by using functions 
$$
\Psi^{[1]}_{j}, \Psi^{[2]}_{j}, \Psi^{[3]}_{j}
$$
in a group $\Pi_{j}, j = 1,...,l-1$ of levels
$$
3(j - 1) + 1, \ 3(j - 1) + 2, \ 3(j - 1) + 3,
$$
i.e., the characterization is the same except it is given in terms
of $\varepsilon_{l}$ and $\delta_{l}$ rather than in terms of
$\varepsilon_{j}$ and $\delta_{j}$, while the parameters are
connected by the renormalizations
$$
\varepsilon_{j+1} = 2^{3}\varepsilon_{j}, \ \delta_{j+1} =
\varepsilon^{3}_{j}\delta_{j}, \ j = 1,...,l-1,
$$
where $\varepsilon_{1} = \varepsilon, \ \delta_{1} = \delta$.

In the group $\Pi_{1}$ at level $1$ the process forms the parts
$$
{\cal P}_{11}^{\eta_{1}} = (P_{10}^{\eta_{1}} \leftrightarrow
P_{20}^{\eta_{2}})^{\eta_{1}},...,
$$
$$
{\cal P}_{N/2,1}^{\eta_{N/2}} = (P_{N-1,0}^{\eta_{N-1}}
\leftrightarrow P_{N0}^{\eta_{N}})^{\eta_{N/2}},
$$
where an elementary part $P_{j0}^{\eta_{j}}, \ j = 1,...,N$ at level
$0$ can be positive if $\eta_{j}= 1$ or negative if $\eta_{j}= -1$
and the symbol $\leftrightarrow$ means that the elementary parts are
connected through a corresponding prime integer relation.

A part ${\cal P}_{j1}^{\eta_{j}}, j = 1,...,N/2$ is defined as a
basic element, when the boundary curve of its geometrical pattern is
given by the function (Figure 3)
$$
\Psi^{[1]}_{1}(t + 2(j-1)\varepsilon_{1})  =
\eta_{j}\Psi^{[1]}_{1}(t), \ \ \ 0 \leq t \leq 2\varepsilon_{1},
$$
where
$$
\Psi^{[1]}_{1}(t) = \left\{
\begin{array}{cl}
\delta_{1}t, \ 0 \leq t \leq \varepsilon_{1}\\
- t + 2\varepsilon_{1}, \  \varepsilon_{1} \leq t \leq
2\varepsilon_{1}.
\end{array}
\right.
$$

In a group $\Pi_{l}$ at level $3(l-1) + 1, \ l = 2,...,p$ the parts
$$
{\cal P}_{1,3(l-1)+1}^{\eta_{1}},..., {\cal
P}_{N/2^{3(l-1)+1},3(l-1)+1}^{\eta_{N/2^{3(l-1)+1}}}
$$
formed by the process from the parts as basic elements at level
$3(l-2) + 1$ in their turn can be defined as basic elements at level
$3(l-1) + 1$, when the boundary curve of the geometrical pattern of
a part
$$
{\cal P}_{j,3(l-1)+1}^{\eta_{j}}, j = 1,...,N/2^{3(l-1)+1}
$$
become specified by the function
$$
\Psi^{[1]}_{l}(t + 2(j-1)\varepsilon_{l})  =
\eta_{j}\Psi^{[1]}_{l}(t), \ \ \ 0 \leq t \leq 2\varepsilon_{l},
$$
where
$$
\Psi^{[1]}_{l}(t) = \left\{
\begin{array}{cl}
\delta_{l}t, \ 0 \leq t \leq \varepsilon_{l}\\
- t + 2\varepsilon_{l}, \  \varepsilon_{l} \leq t \leq
2\varepsilon_{l}
\end{array}
\right.
$$
and the part ${\cal P}_{j,3(l-1)+1}^{\eta_{j}}$ is positive if
$\eta_{j}= 1$ or negative if $\eta_{j}= -1$.

At this step in the construction of the representation the
renormalization
$$
\varepsilon_{l} = 2^{3}\varepsilon_{l-1}, \ \ \ \delta_{l} =
\varepsilon^{3}_{l-1}\delta_{l-1}
$$
introduces new length scales, while the parts, each made of $8$
basic elements of level $3(l-2)+1$, are coarse-grained and become
basic elements of level $3(l-1)+1$.

Second, although in a group $\Pi_{l}$ the parts at level
$$
3(l-1) + q, \ l = 1,...,p, \ q = 1, 2, 3
$$
can be viewed differently as encoded by the functions
$$
\Psi^{[q]}_{l}, \Psi^{[3+q]}_{l-1},..., \Psi^{[3(l-1) + q]}_{1},
$$
yet, because the functions define the geometrical patterns with the
same area
$$
\int_{t_{(j-1)2^{3(l-1) + q}}}^{t_{j2^{3(l-1) + q}}}
\Psi^{[q]}_{l}(t)dt
$$
$$
= \int_{t_{(j-1)2^{3(l-1) + q}}}^{t_{j2^{3(l-1) + q}}} \Psi^{[3 +
q]}_{l-1}(t)dt = ...
$$
$$
= \int_{t_{(j-1)2^{3(l-1) + q}}}^{t_{j2^{3(l-1) + q}}} \Psi^{[3(l-1)
+ q]}_{1}(t)dt,
$$
\begin{equation}
\label{SectionIV.1} j = 1,...,2^{3p+1}/2^{3(l-1) + q}
\end{equation}
as well as the base and height, the parts can be characterized in
the same manner. However, the lengths of the boundary curves of the
geometrical patterns are different.

Therefore, in the representation, irrespective of the basic elements
the part is made of, the energy of the part is conserved.

For example, Figure 3 shows that in the group $\Pi_{2}$ at level $4$
the function $\Psi^{[1]}_{2}$ gives the first term of the series of
approximations in the characterization of the process. The function
$\Psi^{[4]}_{1}$ gives the second and the last term in the finer
resolution, where the basic elements of level $4$ can be seen as the
objects composed from the basic elements of level $1$. Notably, the
geometrical patterns of the first and second terms of the series
have the same area, base and height.

The characterization by the function $\Psi^{[1]}_{l}, l = 2,...,p$
is much simpler than those by the functions $\Psi^{[4]}_{l-1},...,
\Psi^{[3(l-1) + 1]}_{1}$. It assumes that the parts at level $3(l-1)
+ 1$ are basic elements, i.e., have no internal structure, and does
not take into account that they are actually composite objects. At
the same time the simplest characterization can provide the same
information about a number of quantities of the part and, in this
regard, the energy stands out remarkably.

Therefore, in a group $\Pi_{l}, l = 2,...,p$ at level
$$
3(l-1) + q, \ q = 1,2,3
$$
the first term $\Psi^{[q]}_{l}$ of the series can be used as long as
it gives enough information about the part. Moreover, if needed, the
approximation can be improved by taking into account the corrections
from the neglected levels up to the function $\Psi^{[3(l-1) +
q]}_{1}$ providing the information in terms of the basic elements at
level $1$.

It is useful to consider the representation from the reverse
perspective given by the functions
$$
\varphi_{1}(\tau) =
\frac{\Psi^{[1]}_{p}(t)}{\varepsilon_{p}\delta_{p}}, \ \ \
\varphi_{2}(\tau) =
\frac{\Psi^{[4]}_{p-1}(t)}{\varepsilon_{p}\delta_{p}},...,
$$
$$
\varphi_{p}(\tau) = \frac{\Psi^{[3(p-1) +
1]}_{1}(t)}{\varepsilon_{p}\delta_{p}},
$$
$$
\tau = \frac{t}{\varepsilon_{p}}, \ \ \ 0 \leq \tau \leq 2, \ \ \ 0
\leq t \leq 2\varepsilon_{p}, \ \ \ p = 1,2,... \ .
$$

These functions are linearly ordered in the sense that
$$
\varphi_{l+1}(\tau) < \varphi_{l}(\tau), \ \ \ 0 < \tau <
\frac{1}{2}, \ \ \ \frac{3}{2} < \tau < 2,
$$
$$
\varphi_{l+1}(\tau) > \varphi_{l}(\tau), \ \ \ \frac{1}{2} < \tau <
\frac{3}{2}, \ l = 1,...,p-1
$$
and, since
$$
0 \leq \varphi_{l}(\tau) \leq 1, \ \ \ l = 1,...,p, \ \ \ 0 \leq
\tau \leq 2,
$$
can be viewed in the context of a unit circle (Figure 4).

\begin{figure}
\includegraphics[width=.49\textwidth]{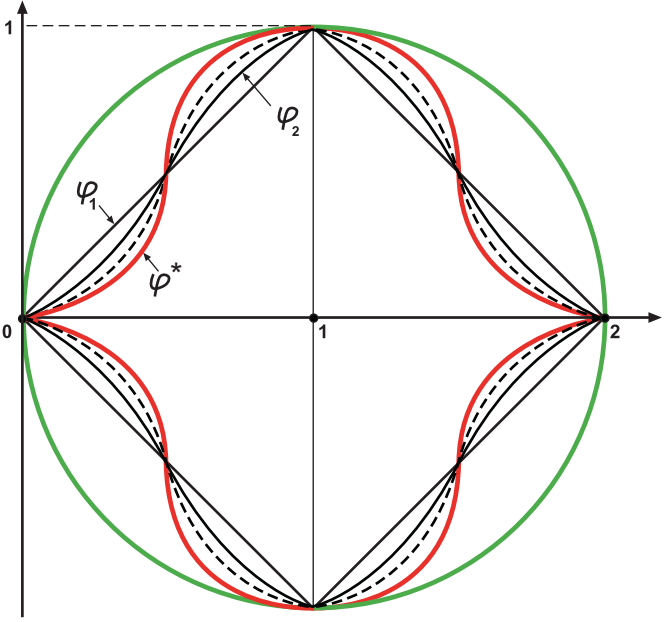}
\caption{\label{fig:two} To help to understand how a unit circle
could be defined in the hierarchical network the figure aims to
represent the sequence of functions (\ref{SectionIV.3}), but without
taking into account the way they may intersect with the unit
circle.}
\end{figure}

From the reverse perspective the conservation of the energy
(\ref{SectionIV.1}) takes the form
$$
E = A = \int_{0}^{2}\varphi_{1}(\tau) d\tau =
\int_{0}^{2}\varphi_{2}(\tau) d\tau \ = ...
$$
\begin{equation}
\label{SectionIV.2} = \int_{0}^{2}\varphi_{l}(\tau) d\tau = 1,
\end{equation}
where $A$ is the area under each of the functions.

By comparing the perspectives we can note that, while the transition
from the function $\Psi^{[4]}_{1}(t), \ 0 \leq t \leq
2\varepsilon_{2}$ to the function $\Psi^{[1]}_{2}(t), \ 0 \leq t
\leq 2\varepsilon_{2}$ hides the information about the hierarchical
structure of the part and thus makes it as a basic element, the
transition from the function $\varphi_{1}, \ 0 \leq \tau \leq 2$ to
the function $\varphi_{2}, \ 0 \leq \tau \leq 2$, however, reveals
the hierarchical structure of the basic element. In particular, as
the function $\varphi_{1}$ specifies the part as a basic element,
the function $\varphi_{2}$ encodes the same part, but as a
hierarchical structure made by the process from $8$ basic elements
of level $1$.

In general, the sequence of functions
\begin{equation}
\label{SectionIV.3} \varphi_{1}, \varphi_{2},..., \varphi_{p}, \ \ \
p = 1,2,...
\end{equation}
allows to consider the part at gradually smaller length scales. And
while the function $\varphi_{1}$ specifies the part as a basic
element, a function $\varphi_{l}, \ l = 2,...,p$ can reveal the part
at a finer resolution, where it is built by the process from
$2^{3l}$ basic elements. Moreover, the transition from a function
$\varphi_{l}, \ l = 1,...,p-1$ to the function $\varphi_{l+1}$
reveals that the supposed to be basic elements are, in fact,
divisible and have their own internal structure. Consequently, it
turns out that the part is made by the process from $2^{3(l+1)}$
basic elements three more levels down.

However, it is interesting to know whether at least in the limit of
the reductions
\begin{equation}
\label{SectionIV.4} \lim_{p \to \infty} \varphi_{p} = \varphi^{*}
\end{equation}
some basic elements not divisible any further can exist. In this
regard, these reductions are quite remarkable. Indeed, although each
next step probes the part at smaller length scales and its
specification becomes more complex, yet, with each reduction the
character of the basic elements remains the same thus suggesting
that, in this sense, they are indivisible.

In the description a prime integer relation is a coherent system
with rigid control of its components, but, at the same time, a very
fragile one. As a prime integer relation is sensitive, so is the
boundary curve of its geometrical pattern. This allows us to
characterize the sequence of functions (\ref{SectionIV.3}) by a
fine-tuned parameter.

Namely, let $L_{l}, \ l = 1,...,p$ be the length of the boundary
curve of the geometrical pattern defined by the graph of the
function $\varphi_{l}$. The length $L_{l}$ of the boundary curve has
an important property. In particular, because the exact value of
$L_{l}$ corresponds to the prime integer relation, it becomes so
sensitive that can not be changed even slightly. Whatever a small
change of the value of $L_{l}$ can be, this will loose the
correspondence with the prime integer relation. Therefore, the
length $L_{l}$ has to be fine-tuned, unless its value is set exactly
right, the relationships are not in place for the part to exist.

In view of this link and condition (\ref{SectionIV.2}) a connection
between the area of a geometrical pattern, as the energy of a
corresponding part, and the length of the boundary curve may encode
the distribution of the energy among the elementary parts. Now, let
us specify such a connection.

In particular, by using functions $\varphi_{l}$ and $-\varphi_{l}, \
l = 1,...,p$ we can define a closed curve ${\mit \Gamma}_{l}$ and
consider it in the context of the unit circle with the center at
$(1,0)$ (Figure 4). The curves ${\mit \Gamma}_{1},...,{\mit
\Gamma}_{p}$ in the limit of (\ref{SectionIV.4}) can define a closed
curve ${\mit \Gamma}^{*}$, which looks like a deformed circle.

By analogy with the ratio of a circle's area $\pi r^{2}$ to its
squared circumference $4 \pi^{2} r^{2}$, a parameter $a_{l}, l =
1,...,p$ can be defined through the closed curve ${\mit \Gamma}_{l}$
to get a connection between the area of the geometrical pattern and
the length of its boundary curve
$$
2A = 2\pi a_{l}(2L_{l})^{2},
$$
where $2A$ is the area enclosed by the curve ${\mit \Gamma}_{l}$,
$2L_{l}$ is the length of the curve and $2\pi$ is a normalization
factor. Therefore, the parameter
$$
a_{l} = \frac{2A_{l}}{2\pi (2L_{l})^{2}} = \frac{A}{4\pi L_{l}^{2}}
$$
relates the area of the geometrical pattern and the length of the
boundary curve, where $A_{l} = A = 1$. Due to the character of
$L_{l}$ the parameter $a_{l}$ is fine-tuned.

We can calculate
$$
a_{1} = \frac{A_{1}}{2\pi L_{1}^{2}} = 0.0099471839 ... \approx
\frac{1}{101},
$$
where $\varepsilon = 1, \ \delta = 1$ (Figure 4).

To find $a_{2}$ we can use the polynomial expressions of the
function $\Psi^{[4]}_{1}(t), \ t_{0} \leq t \leq t_{16}$
$$
\Psi_{1}^{[4]}(t) = \frac{t^{4}}{4!}, \ \ \  t \in [t_{0}, t_{1}],
$$
$$
\Psi_{1}^{[4]}(t) = - \frac{t^{4}}{4!} + \frac{t^{3}}{3} -
\frac{t^{2}}{2} + \frac{t}{3} - \frac{1}{12}, \ \ \ t \in [t_{1},
t_{2}],
$$
$$
\Psi_{1}^{[4]}(t) = - \frac{t^{4}}{4!} + \frac{t^{3}}{3} -
\frac{t^{2}}{2} + \frac{t}{3} - \frac{1}{12}, \ \ \ t \in [t_{2},
t_{3}],
$$
$$
\Psi_{1}^{[4]}(t) = \frac{t^{4}}{4!} - \frac{2t^{3}}{3} + 4t^{2} -
\frac{26t}{3} + \frac{20}{3}, \ \ \ t \in [t_{3}, t_{4}],
$$
$$
\Psi_{1}^{[4]}(t) = - \frac{t^{4}}{4!} + \frac{2t^{3}}{3} - 4t^{2} +
\frac{38t}{3} - \frac{44}{3}, \ \ \ t \in [t_{4}, t_{5}],
$$
$$
\Psi_{1}^{[4]}(t) = \frac{t^{4}}{4!} - t^{3} + \frac{17t^{2}}{2} -
\frac{87t}{3} + \frac{449}{12}, \ \ \ t \in [t_{5}, t_{6}],
$$
$$
\Psi_{1}^{[4]}(t) = \frac{t^{4}}{4!} - t^{3} + \frac{17t^{2}}{2} -
\frac{87t}{3} + \frac{449}{12}, \ \ \ t \in [t_{6}, t_{7}],
$$
$$
\Psi_{1}^{[4]}(t) = - \frac{t^{4}}{4!} + \frac{4t^{3}}{3} - 16t^{2}
+ \frac{256t}{3} - \frac{488}{3}, \ \ \ t \in [t_{7}, t_{8}].
$$
$$
\Psi_{1}^{[4]}(t) = - \frac{t^{4}}{24} + \frac{4t^{3}}{3} - 16t^{2}
+ \frac{256t}{3} - \frac{488}{3}, \ \ \ t \in [t_{8}, t_{9}],
$$
$$
\Psi_{1}^{[4]}(t) = \frac{t^{4}}{24} - \frac{5t^{3}}{3} +
\frac{49t^{2}}{2} - \frac{473t}{3} + \frac{9218}{24}, \ \ \ t \in
[t_{9}, t_{10}],
$$
$$
\Psi_{1}^{[4]}(t) = \frac{t^{4}}{24} - \frac{5t^{3}}{3} +
\frac{49t^{2}}{2} - \frac{473t}{3} + \frac{9218}{24}, \ \ \ t \in
[t_{10}, t_{11}],
$$
$$
\Psi_{1}^{[4]}(t) = - \frac{t^{4}}{24} + 2t^{3} - 36t^{2} +
\frac{858t}{3} - 836, \ \ \ t \in [t_{11}, t_{12}],
$$
$$
\Psi_{1}^{[4]}(t) = \frac{t^{4}}{24} - 2t^{3} + 36t^{2} - 290t +
892, \ \ \ t \in [t_{12}, t_{13}],
$$
$$
\Psi_{1}^{[4]}(t) = - \frac{t^{4}}{24} + \frac{7t^{3}}{3} -
\frac{97t^{2}}{2} + \frac{1327t}{3} - \frac{35714}{24}, \ \ \ t \in
[t_{13}, t_{14}],
$$
$$
\Psi_{1}^{[4]}(t) = - \frac{t^{4}}{24} + \frac{7t^{3}}{3} -
\frac{97t^{2}}{2} + \frac{1327t}{3} - \frac{35714}{24}, \ \ \ t \in
[t_{14}, t_{15}],
$$
$$
\Psi_{1}^{[4]}(t) = \frac{t^{4}}{24} - \frac{8t^{3}}{3} + 64t^{2}
$$
\begin{equation}
\label{SectionIV.5} - \frac{2048t}{3} + \frac{8192}{3}, \ \ \ t \in
[t_{15}, t_{16}]
\end{equation}
to get $\varphi_{2}$ and then by computations obtain
$$
a_{2} = \frac{A_{2}}{2\pi L_{2}^{2}} = 0.0084640979 ... \approx
\frac{1}{118},
$$
where $\varepsilon = 1, \ \delta = 1$.

Although digits in the value of $a_{1}$ and $a_{2}$ seem to appear
at random, yet, each digit there must stand as it is and not be even
a bit different. Remarkably, our description can explain the values
of the parameter, which are uniquely fixed and protected by
arithmetic through the corresponding prime integer relations.

Since the function $\Psi_{1}^{[4]}$ belongs to level $4$, the four
quantities $(\ref{SectionII.2})$ of the complex system remain
unchanged. This invariance can be expressed by the conservation of
four quantum numbers of the elementary parts as the first four
coefficients of their polynomials.

In particular, from (\ref{SectionIV.5}) we can find that the sum of
the first quantum numbers
$$
+ \frac{1}{4!} - \frac{1}{4!} - \frac{1}{4!} + \frac{1}{4!} -
\frac{1}{4!} + \frac{1}{4!} + \frac{1}{4!} - \frac{1}{4!}
$$
$$
- \frac{1}{4!} + \frac{1}{4!} + \frac{1}{4!} - \frac{1}{4!} +
\frac{1}{4!} - \frac{1}{4!} - \frac{1}{4!} + \frac{1}{4!} = 0,
$$
the sum of the second quantum numbers
$$
0 + \frac{1}{3} + \frac{1}{3} - \frac{2}{3} + \frac{2}{3} - 1 - 1 +
\frac{4}{3} + \frac{4}{3}
$$
$$
- \frac{5}{3} - \frac{5}{3} + 2 - 2 + \frac{7}{3} + \frac{7}{3} -
\frac{8}{3} = 0,
$$
the sum of the third quantum numbers
$$
0 - \frac{1}{2} - \frac{1}{2} + 4 - 4 + \frac{17}{2} + \frac{17}{2}
- 16 - 16
$$
$$
+ \frac{49}{2} + \frac{49}{2} - 36 + 36 - \frac{97}{2} -
\frac{97}{2} + 64 = 0
$$
and the sum of the fourth quantum numbers
$$
0 + \frac{1}{3} + \frac{1}{3} - \frac{26}{3} + \frac{38}{3} -
\frac{87}{3} - \frac{87}{3} + \frac{256}{3} + \frac{256}{3} -
\frac{473}{3}
$$
$$
- \frac{473}{3} + \frac{858}{3} - 290 + \frac{1327}{3} +
\frac{1327}{3} - \frac{2048}{3} = 0.
$$
Therefore, the quantum numbers are all preserved.

\section{Transformation of Laws of Arithmetic into Laws of Space and Time}

In the previous section we have obtained a first resolution picture
of the hierarchical network, where the correlation structure
determined by the process at levels $1, 2$ and $3$ is isomorphically
represented by a hierarchical structure of two-dimensional
geometrical patterns.

In this section we consider the transformation of the laws of
arithmetic the elementary parts of the correlation structure are
determined by in the hierarchical network into the laws of the
elementary parts in space and time. Significantly, the laws of
arithmetic can be fully expressed by the hierarchical structure of
geometrical patterns and for an elementary part are entirely given
by its geometrical pattern \cite{Korotkikh_1}-\cite{Korotkikh_6}.

The geometrical pattern of an elementary part has two defining
entities, i.e., the boundary curve and the area. In the
transformation we represent the boundary curve by the space and time
coordinates of the elementary part and the area by the energy of the
elementary part \cite{Korotkikh_6}-\cite{Korotkikh_9}.

Now, let us consider how the boundary curves of the elementary parts
can be represented by using space and time as dynamical variables
(Figure 3).

At level $0$ the space and time coordinates of an elementary part
$P_{j0}, j = 1,...,16$ are defined through the arithmetical and
geometrical forms of the description. In the arithmetical form at
level $0$, where the integers have no relationships, nothing acts on
the elementary part $P_{j0}$ and, therefore, it can be defined in
the state of rest. In the geometrical form, the boundary curve of
the elementary part $P_{j0}$ is given by the piecewise constant
function
$$
\Psi^{[0]}_{1}(t), \ t_{j-1} \leq t < t_{j}
$$
and can be represented by the space and time coordinates of the
elementary part $P_{j0}$ as follows. Namely, at the moment
$T_{j0}(t_{j-1}) = 0$ of the local time, in the transition from one
state into another, the space coordinate $X_{j0}(t_{j-1})$ of the
elementary part $P_{j0}$ changes by
$$
\Delta X_{j0} = \Psi^{[0]}_{1}(t_{j-1}) = \eta_{j}\delta
$$
and then stays as it is, while the time coordinate
$$
T_{j0}(t), \ t_{j-1} \leq t < t_{j},
$$
changes independently with the length of the boundary curve
$$
\Delta T_{j0}(t) = T_{j0}(t) - T_{j0}(t_{j-1}) = T_{j0}(t)
$$
$$
= \int_{t_{j-1}}^{t}\sqrt{1 + (\frac{d\Psi^{[0]}_{1}}{dt})^{2}}dt,
$$
where we set
$$
\Delta T_{j0} = \lim_{t \to t_{j}} \int_{t_{j-1}}^{t}\sqrt{1 +
(\frac{d\Psi^{[0]}_{1}}{dt})^{2}}dt = \varepsilon.
$$
Notably, no matter what the space coordinate is, the time coordinate
is not affected.

Figure 3 shows that under the integration of the function
$\Psi^{[l]}_{1}(t), \ l = 0,1,2, \ t_{0} \leq t \leq t_{16}$ the
geometrical patterns of the elementary parts at level $l$ transform
into the geometrical patterns of the elementary parts at level
$l+1$. As a result, the boundary curve of an elementary part
$P_{jl}, j = 1,...,16$, i.e., the graph of the function
$$
\Psi^{[l]}_{1}(t), \ t_{j-1} \leq t \leq t_{j},
$$
transforms into the boundary curve of the elementary part
$P_{j,l+1}$, i.e., the graph of the function
$$\Psi^{[l+1]}_{1}(t), \ t_{j-1} \leq t \leq t_{j}.
$$

Significantly, under the transformations of the geometrical patterns
space and time become dynamic variables defined by the
representation of the boundary curves.

In particular, at level $1$ the space coordinate $X_{j1}(t)$ and
time coordinate $T_{j1}(t), t_{j-1} \leq t \leq t_{j}$ of an
elementary part $P_{j1}, j = 1,...,16$ become linearly dependent and
characterize the motion of the elementary part $P_{j1}$ by
\begin{equation}
\label{SectionV.1} \Delta T_{j1}(t) sin(\alpha_{j}) = \Delta
X_{j1}(t),
\end{equation}
with
$$
\Delta X_{j1}(t) = X_{j1}(t) - X_{j1}(t_{j-1})
$$
\begin{equation}
\label{SectionV.2} = \Psi^{[1]}_{1}(t) - \Psi^{[1]}_{1}(t_{j-1})
\end{equation}
and
$$
\Delta T_{j1}(t) = \int_{t_{j-1}}^{t}\sqrt{1 +
(\frac{d\Psi^{[1]}_{1}(t')}{dt'})^{2}}dt'
$$
$$
= \int_{t_{j-1}}^{t}\sqrt{1 + (\frac{dX_{j1}(t')}{dt'})^{2}}dt',
$$
where the angle $\alpha_{j}$ is given by
$$
tan (\alpha_{j}) = \Psi^{[0]}_{1}(t_{j-1}).
$$

Let
$$
\Delta X_{j1} =  X_{j1}(t_{j}) - X_{j1}(t_{j-1})
$$
and, since $T_{j1}(t_{j-1}) = 0$,
$$
\Delta T_{j1} = T_{j1}(t_{j}) - T_{j1}(t_{j-1}) = T_{j1}(t_{j}).
$$

The velocity $V_{j1}(t), \ t_{j-1} \leq t \leq t_{j}$ of the
elementary part $P_{j1}$, as a dimensionless quantity, can be
defined by
\begin{equation}
\label{SectionV.3} V_{j1}(t) = \frac{\Delta X_{j1}(t)}{\Delta
T_{j1}(t)}.
\end{equation}
Applying conditions $(\ref{SectionV.1})$ and $(\ref{SectionV.3})$,
we obtain
$$
V_{j1}(t) = sin (\alpha_{j})
$$
and, since the angle $\alpha_{j}$ is constant, the velocity is also
a constant
$$
V_{j1}(t) = V_{j1}.
$$

By definition $-1 \leq sin (a_{j}) \leq 1$ and so
\begin{equation}
\label{SectionV.4} -1 \leq  V_{j1} \leq 1.
\end{equation}

Since the velocity $V_{j1}$ is a dimensionless quantity, the
condition $(\ref{SectionV.4})$ determines a velocity limit, which we
associate with the speed of light $c$. We can now define the
dimensional velocity $v_{j1}$ of the elementary part $P_{j1}$ by
\begin{equation}
\label{SectionV.5} V_{j1} = sin(\alpha_{j}) = \frac{v_{j1}}{c}
\end{equation}
and, therefore, $ \vert v_{j1} \vert \leq c$.

By taking into account that at level $0$
$$
\vert \Delta X_{j0} \vert =\vert \eta_{j} \delta \vert = \vert \pm
\delta \vert = \delta, \ \ \ \Delta T_{j0} = \varepsilon,
$$
from the geometry of the process at level $1$, we can find the
dimensionless quantities
$$
\frac{\chi_{j1}}{\chi_{min,j1}}, \ \ \
\frac{\tau_{j1}}{\tau_{min,j1}}
$$
of space and time of the elementary part $P_{j1}$.

In particular, for the dimensionless quantity
$$
\frac{\chi_{j1}}{\chi_{min,j1}}
$$
of space, by using $(\ref{SectionV.2})$, we obtain
$$
\frac{\chi_{j1}}{\chi_{min,j1}} = \vert X_{j1}(t_{j}) -
X_{j1}(t_{j-1}) \vert
$$
\begin{equation}
\label{SectionV.6}  = \vert \Psi^{[1]}_{1}(t_{j}) -
\Psi^{[1]}_{1}(t_{j-1}) \vert = \delta \varepsilon =
\frac{\chi_{0}}{\chi_{min}}\frac{\tau_{0}}{\tau_{min}}.
\end{equation}

Therefore, the absolute change of the dimensional space coordinate
of the elementary part $P_{j1}$ is $\chi_{0}\tau_{0}$. It sets the
length scale of space at level $1$ as
$$
\chi_{1} = \chi_{j1} = \chi_{0}\tau_{0},
$$
while the minimum length scale of space $\chi_{min,1}$ at level $1$
is given by
$$
\chi_{min,1} = \chi_{min,j1} = \chi_{min}\tau_{min}.
$$

To find the dimensionless quantity
$$
\frac{\tau_{j1}}{\tau_{min,j1}}
$$
of time of the elementary part $P_{j1}$ we can use the Pythagorean
theorem
$$
\frac{\tau_{j1}^{2}}{\tau_{min,1}^{2}} = \varepsilon^{2} +
\delta^{2}\varepsilon^{2} = \frac{\tau^{2}_{0}}{\tau_{min}^{2}}
+
\frac{\chi^{2}_{0}}{\chi_{min}^{2}}\frac{\tau^{2}_{0}}{\tau_{min}^{2}}
$$
$$
=
\frac{\chi_{min}^{2}}{\chi_{min}^{2}}\frac{\tau^{2}_{0}}{\tau_{min}^{2}}
+
\frac{\chi^{2}_{0}}{\chi_{min}^{2}}\frac{\tau^{2}_{0}}{\tau_{min}^{2}}
$$
and find
\begin{equation}
\label{SectionV.7} \frac{\tau_{j1}}{\tau_{min,1}} = \frac{\tau_{0}
\sqrt{\chi_{min}^{2} + \chi^{2}_{0}}}{\chi_{min}\tau_{min}}.
\end{equation}

According to $(\ref{SectionV.7})$, the change of the dimensional
time coordinate of the elementary part $P_{j1}$ is
$$
\tau_{1} = \tau_{j1} = \tau_{0} \sqrt{\chi_{min}^{2} +
\chi^{2}_{0}}.
$$
It sets the length scale of time at level $1$, while the minimum
length scale of time at level $1$ is given by
$$
\tau_{min,1} = \tau_{min,j1} = \chi_{min}\tau_{min}.
$$

Importantly, by using conditions $(\ref{SectionV.6})$ and
$(\ref{SectionV.7})$ we can express $\frac{v_{j1}}{c}$ as follows
\begin{equation}
\label{SectionV.8} \vert \frac{v_{j1}}{c} \vert = \vert
sin(\alpha_{j}) \vert =
\frac{\frac{\chi_{0}}{\chi_{min}}\frac{\tau_{0}}{\tau_{min}}}
{\frac{\tau_{0} \sqrt{\chi_{min}^{2} +
\chi^{2}_{0}}}{\chi_{min}\tau_{min}}} =
\frac{\chi_{0}}{\sqrt{\chi_{min}^{2} + \chi^{2}_{0}}}.
\end{equation}

Therefore, the minimum length scale of space $\chi_{min}
> 0$ determines that the absolute value of the velocity $v_{j1}$ can
not be equal to the velocity limit $c$. Indeed,
$$
\vert \frac{v_{j1}}{c} \vert = \frac{\chi_{0}}{\sqrt{\chi_{min}^{2}
+ \chi^{2}_{0}}} < 1
$$
and thus
$$
\vert v_{j1} \vert < c.
$$
However, when $\chi_{min} = 0$ or $\chi_{min} << \chi_{0}$ the
velocity $v_{j1}$ can become equal
\begin{equation}
\label{SectionV.9} \vert v_{j1} \vert = c
\end{equation}
or very close to $c$.

The condition $(\ref{SectionV.9})$ suggests a parallel between the
elementary parts of level $1$ and photons traveling at the speed of
light. This actually motivated us to use the notation $c$ for the
velocity limit and associate it with the speed of light.

Notably, when $\chi_{0} = \chi_{min}$, then
$$
\vert \frac{v_{j1}}{c} \vert = \frac{\chi_{0}}{\sqrt{\chi_{min}^{2}
+ \chi^{2}_{0}}}
$$
$$
= \frac{\chi_{min}}{\sqrt{\chi_{min}^{2} + \chi_{min}^{2}}} =
\frac{1}{\sqrt{2}}
$$
and thus for the velocity $v_{j1}$ of the elementary part $P_{j1}$
we obtain
$$
\vert v_{j1} \vert = \frac{c}{\sqrt{2}}.
$$
As a result, we might say that when $\chi_{0} = \chi_{min}$ photons
travel slower than in the case of $(\ref{SectionV.9})$.

Now, let us consider how the times $\Delta T_{j0}$ and $\Delta
T_{j1}$ of the elementary parts $P_{j0}$ and $P_{j1}, j = 1,...,16$
can be connected. Note that, while there are no relationships to
affect the elementary part $P_{j0}$ and thus it remains in the state
of rest, the elementary part $P_{j1}$ is forced to move as a result
of the relationship with another elementary part.

Figure 3 shows that
$$
\Delta T_{j1} cos(\alpha_{j}) = \Delta T_{j0}
$$
and, by using $(\ref{SectionV.5})$, we get
\begin{equation}
\label{SectionV.10} \Delta T_{j1} = \frac{\Delta T_{j0}}{\sqrt{ 1 -
\frac{v_{j1}^{2}}{c^{2}}}}.
\end{equation}

Since under the prime integer relations the correlations are
realized simultaneously, then from $(\ref{SectionV.10})$ we can find
that the time $T_{j1}(t)$ of the elementary part $P_{j1}$ runs
faster than the time $T_{j0}(t), \ t_{j-1} \leq t \leq t_{j}$ of the
elementary part $P_{j0}$.

Remarkably, the condition $(\ref{SectionV.10})$ symbolically
reproduces the well-known formula connecting the elapsed times in
the moving and the stationary systems \cite{Einstein_2} and allows
its interpretation. In particular, as long as from a common
perspective one tick of the clock of the moving elementary part
$P_{j1}$ takes longer $\Delta T_{j1} > \Delta T_{j0}$ than one tick
of the clock of the stationary elementary part $P_{j0}$, then the
time counted by the number of ticks of the clock in the moving
system will be less than the time counted by the number of ticks of
the clock in the stationary system.

Notably, at level $1$ the motion of the elementary part $P_{j1}$ has
the invariant
\begin{equation}
\label{SectionV.11} \Delta T_{j1}^{2} - \Delta X_{j1}^{2} =
\varepsilon^{2}
\end{equation}
with recognizable features of the Lorentz invariant.

Significantly, starting with level $2$, in the representation of the
boundary curve the space and time coordinates of an elementary part
$P_{jl}, j = 1,...,16, \ l = 2,3$ become intimately linked, so that
the boundary curve can be seen as their joint entity we define as
the local spacetime of the elementary part $P_{jl}$. As for the
boundary curves of the elementary parts at levels $0$ and $1$, we
also define them as their local spacetimes.

In particular, in the representation of the boundary curve given by
the graph of the function
$$
\Psi^{[l]}_{1}(t), \ t_{j-1} \leq t \leq t_{j}, \ j =
1,...,16, \ l = 2,3
$$
the space coordinate $X_{jl}(t)$ of the elementary part $P_{jl}$ is
defined by
\begin{equation}
\label{SectionV.12} X_{jl}(t) = \Psi^{[l]}_{1}(t), \ \ \ t_{j-1}
\leq t \leq t_{j}.
\end{equation}
In its turn the time coordinate $T_{jl}(t)$ of the elementary part 
$P_{jl}$ is defined by the length of the curve
$$
T_{jl}(t) = \int_{t_{j-1}}^{t}\sqrt{1 +
(\frac{d\Psi^{[l]}_{1}(t')}{dt'})^{2}}dt'
$$
\begin{equation}
\label{SectionV.13} = \int_{t_{j-1}}^{t}\sqrt{1 +
(\frac{dX_{jl}(t')}{dt'})^{2}}dt', \ \ \ t_{j-1} \leq t \leq t_{j}
\end{equation}
and, as a result, the space and time coordinates become
interdependent.

In the representation the motion of the elementary part $P_{jl}$ can
be defined by the change of the space coordinate $X_{jl}(t)$ with
respect to time coordinate $T_{jl}(t)$. Namely, as the time
coordinate $T_{jl}(t)$ changes by
$$
\Delta T_{jl}(t) = \int_{t_{j-1}}^{t}\sqrt{1 +
(\frac{d\Psi^{[l]}_{1}(t)}{dt})^{2}}dt
$$
$$
= \int_{t_{j-1}}^{t}\sqrt{1 + (\frac{dX_{jl}(t)}{dt})^{2}}dt
$$
the space coordinate $X_{jl}(t)$ changes by
$$
\Delta X_{jl}(t) = \Psi^{[l]}_{1}(t) - \Psi^{[l]}_{1}(t_{j-1})
$$
and we may say that under the correlations the position of the
elementary part $P_{jl}$ during the time
$$
\Delta T_{jl} = \int_{t_{j-1}}^{t_{j}}\sqrt{1 +
(\frac{d\Psi^{[l]}_{1}(t)}{dt})^{2}}dt
$$
$$
= \int_{t_{j-1}}^{t_{j}}\sqrt{1 + (\frac{dX_{jl}(t)}{dt})^{2}}dt
$$
changes by
$$
\Delta X_{jl} = \Psi^{[l]}_{1}(t_{j}) - \Psi^{[l]}_{1}(t_{j-1}).
$$

For example, for the space coordinate $X_{12}(t)$ and time
coordinate $T_{12}(t)$ of the elementary part $P_{12}$ we can get
$$
X_{12}(t) = \frac{t^{2}}{2}, \ \ \ t_{0} \leq t \leq t_{1}
$$
and
$$
T_{12}(t) = \int_{t_{0}}^{t}\sqrt{1 +
(\frac{dX_{12}(t')}{dt'})^{2}}dt'
$$
$$
= \int_{t_{0}}^{t}\sqrt{1 + t'^{2}}dt'
$$
$$
= \frac{t}{2}\sqrt{1 + t^{2}} + \frac{1}{2} ln(t + \sqrt{1 +
t^{2}}), \ \ \ t_{0} \leq t \leq t_{1}.
$$

We can obtain that
$$
(\frac{dT_{12}(t)}{dt})^{2} - (\frac{dX_{12}(t)}{dt})^{2} = 1.
$$
Indeed,
$$
\frac{dT_{12}(t)}{dt} = \frac{d}{dt}(\frac{t}{2}\sqrt{1 + t^{2}} +
\frac{1}{2} ln(t + \sqrt{1 + t^{2}}))
$$
$$
= \frac{\sqrt{1 + t^{2}}}{2} + \frac{t^{2}}{2\sqrt{1 + t^{2}}} +
\frac{1 + \frac{t}{\sqrt{1 + t^{2}}}}{2(t + \sqrt{1 + t^{2}})}
$$
$$
= \frac{1 + t^{2} + t^{2}}{2\sqrt{1 + t^{2}}} + \frac{\sqrt{1 +
t^{2}} + t}{2(\sqrt{1 + t^{2}})(t + \sqrt{1 + t^{2}})}
$$
$$
= \sqrt{1 + t^{2}}
$$
and
$$
\frac{dX_{12}(t)}{dt} = t,
$$
so
$$
(\frac{dT_{12}(t)}{dt})^{2} - (\frac{dX_{12}(t)}{dt})^{2} = 1 +
t^{2} - t^{2} = 1.
$$

In general, by using $(\ref{SectionV.13})$, we can find that the
motion of an elementary part $P_{jl}, j = 1,...,16, \ l = 2,3$ has
the following invariant
\begin{equation}
\label{SectionV.15} (\frac{dT_{jl}(t)}{dt})^{2} -
(\frac{dX_{jl}(t)}{dt})^{2} = 1
\end{equation}
with the invariant $(\ref{SectionV.11})$ as a special case.

Notably, similar to level $1$, at levels $2$ and $3$ we can get linear
approximations to the dynamics of the elementary parts that provide
a Lorentz type invariant
\begin{equation}
\label{SectionV.16} \Delta \breve T_{jl}^{2} - \Delta \breve
X_{jl}^{2} = \varepsilon^{2},
\end{equation}
while preserving the sum of the energies of the elementary parts.

For example, at level $3$ the approximation for the part
${\cal P}_{13}^{+}$ is given by the functions
$$
y = \frac{2\delta\varepsilon^{3}}{t_{4}}t, \ \ \ t_{0} \leq t \leq
t_{4},
$$
$$
y = - \frac{2\delta\varepsilon^{3}}{t_{8} - t_{4}}t +
\frac{2\delta\varepsilon^{3}}{t_{8} - t_{4}}t_{8}, \ \ \ t_{4} \leq
t \leq t_{8}
$$
and at level $4$ the approximation for the part ${\cal P}_{14}^{+}$
is specified by the function $\Psi^{[1]}_{2}(t), \ t_{0} \leq t \leq
t_{16}$ (Figure 3).

Next, we consider the connection between the local spacetime and the
energy of an elementary part $P_{jl}, j = 1,...,16, \ l = 0,1,2,3$
determined by the representation of the geometrical pattern. In
particular, we can see that the boundary curve, i.e., the graph of the 
function $\Psi^{[l]}_{1}(t), \ t_{j-1} \leq t \leq t_{j}$, not only 
encodes the local spacetime, but also determines the energy of the 
elementary part $P_{jl}$
$$
E_{jl} = \int_{t_{j-1}}^{t_{j}}\vert \Psi^{[l]}_{1}(t)\vert dt.
$$

We can also define the energy density ${\cal E}_{jl}(t)$ of the
elementary part $P_{jl}$
$$
E_{jl} = \int_{t_{j-1}}^{t_{j}}{\cal E}_{jl}(t)dt =
\int_{t_{j-1}}^{t_{j}}\vert \Psi^{[l]}_{1}(t)\vert dt
$$
and, since (\ref{SectionV.12}), obtain
\begin{equation}
\label{SectionV.17} {\cal E}_{jl}(t) = \vert X_{jl}(t) \vert, \ \ \
t_{j-1} \leq t \leq t_{j}
\end{equation}
meaning that the energy density of the elementary part equals its
space coordinate.

To consider the connection between the local spacetimes and the
energies of elementary parts at different levels the energy profile
$E_{jl}(t)$ of an elementary part $P_{jl}, j = 1,...,16, \ l =
0,1,2$
$$
E_{jl}(t) = \int_{t_{j-1}}^{t} \vert \Psi^{[l]}_{1}(t') \vert dt', \
\ \ t_{j-1} \leq t \leq t_{j}
$$
can be useful. Indeed, by using the fundamental theorem of calculus
we can find that the energy profile of the elementary part $P_{jl}$
at level $l$ determines the local spacetime of the elementary part
$P_{j,l+1}$ at level $l+1$
$$
X_{j,l+1}(t) = \Psi^{[l+1]}_{1}(t) =
\int_{t_{0}}^{t}\Psi^{[l]}_{1}(t')dt'
$$
$$
= \int_{t_{0}}^{t_{j-1}}\Psi^{[l]}_{1}(t')dt' +
\int_{t_{j-1}}^{t}\Psi^{[l]}_{1}(t')dt'
$$
$$
= X_{j,l+1}(t_{j-1}) \pm E_{jl}(t), \ \ \ t_{j-1} \leq t \leq t_{j}.
$$

In particular, when $t = t_{j}$, we get
$$
\vert X_{j,l+1}(t_{j}) - X_{j,l+1}(t_{j-1}) \vert = E_{jl}
$$
and, therefore, the energy of the elementary part $P_{jl}$ at level
$l$ is equal to the change of the space coordinate of the elementary
part $P_{j,l+1}$ at level $l+1$.

As the laws of the elementary parts in terms of arithmetic have been
transformed into the laws of the elementary parts in terms of space
and time, now let us consider the resulting structure of the local
spacetimes.

Remarkably, Figure 3 shows the local spacetimes of the elementary
parts all at once and helps to illustrate the notion of simultaneity
in terms of the local spacetimes. Namely, as the correlation
structure turns to be operational, then through the prime integer
relations the elementary parts, irrespective of the distances and
levels, all become instantaneously connected and move
simultaneously, so that the local spacetimes can geometrically
reproduce the prime integer relations in control of the correlation
structure. In other words, the local spacetimes all function
together to be mutually self-consistent in reproducing of the prime
integer relations.

The local spacetime changes with the level and takes the shape
precisely as determined by the prime integer relation. In
particular, this defines that the time runs differently at the
levels and we may say that the rate at which the clock ticks varies
with the level of the correlation structure. Furthermore, for
elementary parts of the same level the elapsed times can be also
different. For example, it can be calculated that the space
coordinate of the elementary part $P_{13}$ changes by $\Delta X_{13}
= 1/6$ during the time $\Delta T_{13} \approx 1.02$, while the space
coordinate of the elementary part $P_{23}$ changes by $\Delta X_{23}
= 5/6$ during the time $\Delta T_{23} \approx 1.30$, when $\delta =
1, \ \varepsilon = 1$.

We may visualize the correlations at work by imagining some points
that move along the boundary curves with simultaneous start and
finish. Since the length of the trajectory of the point so far shows
the time of the elementary part, then observing the point moving at
level $l = 2, 3$ might seem like observing the time curving in the
flow.

Notably, when the boundary curve is viewed as the geodesic of the
elementary part, it appears that the elementary part moves from one
point to another not to extremize an action in between, but to
geometrically reproduce the prime integer relation. In other words,
the prime integer relation guides the motion of the elementary part.
Significantly, the geodesics of the elementary parts are elements 
of one and the same structure, i.e, the hierarchical network of 
prime integer relations.

In analogy with general relativity, where spacetime curves in
respond to energy \cite{Einstein_3}, in our description the local
spacetime of an elementary part also curves in accordance with the
energy of the elementary part. Importantly, the dependence between
the local spacetime and the energy appears as a result of the
transformation of a corresponding law of arithmetic. Furthermore,
the condition for an elementary part to be integrated into the
correlation structure is entirely determined by the geometry of the
local spacetime. Therefore, once we interpret that the elementary
parts are held in the correlation structure by a force, then the
force acting on an elementary part can be fully defined by the
geometry of the local spacetime.

In general, Figure 3 gives us a powerful perspective. First, we can
observe how the local spacetimes appear to be related to one
another. Moreover, we can know in advance what will happen to the
elementary parts even before the correlation structure become
triggered. When that happens, the elementary parts are controlled
nonlocally, but act locally for one common purpose to preserve the
system. Second, the picture is timeless in the sense that past,
present and future are all united in one whole still. Third, because
the elementary parts move in their own local spacetimes, then by
changing the focus from one elementary part to another seems like we
travel not only in space, but in time as well.

\section{Global Spacetimes as Effective Representations
of the Hierarchical Network}

In the previous section we have considered a representation of the
hierarchical structure of geometrical patterns by using space and
time as dynamic variables. As a result, the elementary parts become
specified not in one global spacetime, where the laws of arithmetic
they realize can be expressed by using the same number of space and
time variables, but by local spacetimes and energies.

This fact is not surprising. As the process takes place not in space
and time, but in the hierarchical network, rather than to emerge in
a global spacetime, the local spacetimes have to make the
geometrical patterns corresponding to the hierarchical structure of
prime integer relations.

Remarkably, in the representation the elementary parts act as the
carriers of the laws of arithmetic the process is governed by with
each elementary part carrying its own quantum of the laws entirely
determined by the geometrical pattern. This opens an important 
perspective to use elementary parts in the hierarchical network 
as quanta to construct different laws. In this regard, a global 
spacetime could serve as a common stage, where elementary parts would 
be combined with their quanta of the laws taking the form in terms 
of the same number of space and time variables. Significantly, once 
through the form of the laws a desired objective would be realized, 
the global spacetime could be used as an effective representation of 
the process.

In general, this perspective suggests the hierarchical network as a
source of laws that could be harnessed. In particular, for a given
objective the hierarchical network could be used to generate
self-organization processes to obtain relevant laws of arithmetic
and then process them into the required form by constructing a
corresponding global spacetime.

In fact, the perspective allows us to speculate about an observing
system that in the processing of the hierarchical network could be
adaptable to obtain different effective representations.
Furthermore, the description can provide a formal means in the
context of the mind-matter problem \cite{Kant_1}-\cite{Penrose_1} to
consider why the mind in the processing of the hierarchical network
might be programmed to sense three dimensions of space and one
dimension of time.

Significantly, an effective representation can not function, unless
supported by an equivalence class of inertial reference frames,
where the form of the laws is actually represented and thus the
same. This determines a symmetry group of coordinate transformations
with the form as the invariant and resonates well with the principle
of relativity.

Historically, it has been established that physical laws can express
themselves through the same form when considered in their inertial
reference frames and this has resulted in the principle of
relativity \cite{Einstein_2}. For example, in classical mechanics
the Galilean transformations specify the inertial coordinate
systems, while in electromagnetic theory the Lorentz transformations
take responsibility for the transitions between the inertial
reference frames.

However, the principle of relativity and the description
may follow opposite directions. Namely, the principle of
relativity, based on the success with a number of physical laws,
is tried to make a leap forward and accommodate all physical
laws. Clearly, it would be an ideal situation, instead of
discovering laws one by one, to establish and maintain them all as
one source of the physical laws to rely on when needed. For example,
when a problem arises the source could be used to generate specific
laws to solve the problem.

In the description, by contrast, all its possible laws are already
given. They are the laws of arithmetic realized by the processes in
the construction of the hierarchical network. Consequently, the
hierarchical network appears as a source of laws that could be used
to supply particular laws on demand by defining a corresponding
global spacetime.

In this section we consider a number of effective representations of
the process in terms of global spacetimes.

First, we consider a representation, where through the local
spacetimes the elementary parts $P_{j1}, P_{j2}$ and $P_{j3}, j =
1,...,16$ are combined by a global spacetime
$$
{\cal R}_{j}(3,3) = {\cal R}_{j1} \times {\cal R}_{j2} \times {\cal
R}_{j3}
$$
as the direct product of two-dimensional Euclidean spaces ${\cal
R}_{j1}, {\cal R}_{j2}$ and ${\cal R}_{j3}$.

In particular, by using ${\cal R}_{jl}, j = 1,...16, \ l = 1,2,3$
the space and time coordinates of the elementary part $P_{jl}$ are
specified by a vector function
$$
{\bf R}_{jl}(t) = T_{jl}(t){\bf w}_{jl} + \sqrt{-1}X_{jl}(t){\bf
u}_{jl}
$$
\begin{equation}
\label{SectionVI.1} = \sqrt{-1}X_{jl}(t){\bf u}_{jl} + T_{jl}(t){\bf
w}_{jl}, \ \ \ t_{j-1} \leq t \leq t_{j},
\end{equation}
where ${\bf w}_{jl}$ and ${\bf u}_{jl}$ are the unit vectors of an
orthogonal basis and, to preserve the invariant
$(\ref{SectionV.15})$, the space coordinate is treated imaginary
$\sqrt{-1}X_{jl}(t)$ with
$$
(\sqrt{-1})^{2} = -1.
$$
As a result, ${\cal R}_{jl}$ acquires a Minkowski spacetime type of
signature $(-, +)$.

In the global spacetime ${\cal R}_{j}(3,3)$ the vector functions
$$
{\bf R}_{j1}(t), \ {\bf R}_{j2}(t), \ {\bf R}_{j3}(t), \ \ \ t_{j-1}
\leq t \leq t_{j}
$$
define the curves as the trajectories of the elementary parts
$P_{j1}, P_{j2}$ and $P_{j3}$. In particular, the curve given by the
vector function ${\bf R}_{jl}(t), l = 1,2,3$ can encode the space
coordinate $X_{jl}(t)$ as a function of the time coordinate
$T_{jl}(t)$ and thus the dynamics of the elementary part $P_{jl}$ in
${\cal R}_{j}(3,3)$.

Defined by the direct product of ${\cal R}_{j1}, {\cal R}_{j2}$ and
${\cal R}_{j3}$ the six-dimensional global spacetime ${\cal
R}_{j}(3,3)$ of three space and three time variables does not
contain information about the formation and connection
between the elementary parts $P_{j1}, P_{j2}$ and $P_{j3}$. As a
result, in the global spacetime the elementary parts become seen as
separate entities.

Namely, the elementary part $P_{j1}$ is characterized by the vector
function
$$
{\bf R}_{j1}(t) = \sqrt{-1}X_{j1}(t){\bf u}_{j1} + T_{j1}(t){\bf
w}_{j1}
$$
$$
+ \sqrt{-1} \cdot 0{\bf u}_{j2} + 0{\bf w}_{j2} + \sqrt{-1} \cdot
0{\bf u}_{j3} + 0{\bf w}_{j3}
$$
$$
= \sqrt{-1}X_{j1}(t){\bf u}_{j1} + \sqrt{-1} \cdot 0{\bf u}_{j2} +
\sqrt{-1} \cdot 0{\bf u}_{j3}
$$
\begin{equation}
\label{SectionVI.2} + T_{j1}(t){\bf w}_{j1} + 0{\bf w}_{j2} + 0{\bf
w}_{j3}, \ \ \ t_{j-1} \leq t \leq t_{j}
\end{equation}
with the space coordinates
$$
\sqrt{-1}X_{j1}(t), \sqrt{-1} \cdot 0, \sqrt{-1} \cdot 0
$$
and time coordinates
$$
T_{j1}(t), 0, 0.
$$

The elementary part $P_{j2}$ is characterized by the vector function
$$
{\bf R}_{j2}(t) = \sqrt{-1} \cdot 0{\bf u}_{j1} + 0{\bf w}_{j1}
$$
$$
+ \sqrt{-1}X_{j2}(t){\bf u}_{j2} + T_{j2}(t){\bf w}_{j2}
$$
$$
+ \sqrt{-1} \cdot 0{\bf u}_{j3} + 0{\bf w}_{j3}
$$
$$
= \sqrt{-1} \cdot 0{\bf u}_{j1} + \sqrt{-1}X_{j2}(t){\bf u}_{j2} +
\sqrt{-1} \cdot 0{\bf u}_{j3}
$$
\begin{equation}
\label{SectionVI.3} + 0{\bf w}_{j1} + T_{j1}(t){\bf w}_{j2} + 0{\bf
w}_{j3}, \ \ \ t_{j-1} \leq t \leq t_{j}
\end{equation}
with the space coordinates
$$
\sqrt{-1} \cdot 0, \sqrt{-1}X_{j2}(t), \sqrt{-1} \cdot 0
$$
and time coordinates
$$
0, T_{j1}(t), 0.
$$

And, finally, the elementary part $P_{j3}$ is characterized by the
vector function
$$
{\bf R}_{j3}(t) = \sqrt{-1} \cdot 0{\bf u}_{j1} + 0{\bf w}_{j1} +
\sqrt{-1} \cdot 0{\bf u}_{j2} + 0{\bf w}_{j3}
$$
$$
+ \sqrt{-1}X_{j3}(t){\bf u}_{j3} + T_{j3}(t){\bf w}_{j3}
$$
$$
= \sqrt{-1} \cdot 0{\bf u}_{j1} + \sqrt{-1} \cdot 0{\bf u}_{j2} +
\sqrt{-1}X_{j3}(t){\bf u}_{j3}
$$
\begin{equation}
\label{SectionVI.4} + 0{\bf w}_{j1} + 0{\bf w}_{j2} + T_{j1}(t){\bf
w}_{j3}, \ \ \ t_{j-1} \leq t \leq t_{j}
\end{equation}
with the space coordinates
$$
\sqrt{-1} \cdot 0, \sqrt{-1} \cdot 0, \sqrt{-1}X_{j3}(t)
$$
and time coordinates
$$
0, 0, T_{j1}(t).
$$

By construction this reference frame is special and from
$(\ref{SectionVI.2})$-$(\ref{SectionVI.4})$ we can clearly see that
there exist preferred directions in the global spacetime ${\cal
R}_{j}(3,3)$ thus making it anisotropic. However, the anisotropy
might be hidden in a reference frame, where for the laws to have the
same form the space and time coordinates of an elementary part
$P_{jl}, j = 1,...,16, \ l = 1,2,3$ would be transformed into the
space coordinates
$$
\sqrt{-1}X_{jl1}'(t), \sqrt{-1}X_{jl2}'(t), \sqrt{-1}X_{jl3}'(t)
$$
and time coordinates
$$
T_{jl1}'(t), T_{jl2}'(t), T_{jl3}'(t),
$$
but preserving the invariant $(\ref{SectionV.15})$
$$
(\frac{dT_{jl1}'(t)}{dt})^{2} + (\frac{dT_{jl2}'(t)}{dt})^{2} +
(\frac{dT_{jl3}'(t)}{dt})^{2}
$$
$$
- (\frac{dX_{jl1}'(t)}{dt})^{2} - (\frac{dX_{jl2}'(t)}{dt})^{2} -
(\frac{dX_{jl3}'(t)}{dt})^{2}
$$
\begin{equation}
\label{SectionVI.5} = (\frac{dT_{jl}(t)}{dt})^{2} -
(\frac{dX_{jl}(t)}{dt})^{2} = 1, \ \ \ t_{j-1} \leq t \leq t_{j}.
\end{equation}

As a result, the elementary part $P_{jl}$ would be characterized by
a vector function
$$
{\bf R}_{jl}'(t)
$$
$$
= \sqrt{-1}X_{jl1}'(t){\bf u}_{j1}' + \sqrt{-1}X_{jl2}'(t){\bf
u}_{j2}' + \sqrt{-1}X_{jl3}'(t){\bf u}_{j3}'
$$
\begin{equation}
\label{SectionVI.6} + \ T_{jl1}'(t){\bf w}_{j1}' + T_{jl2}'(t){\bf
w}_{j2}' + T_{jl3}'(t){\bf w}_{j3}',
\end{equation}
where $t_{j-1} \leq t \leq t_{j}$. Therefore, the invariant
$(\ref{SectionVI.5})$ would define an equivalence class of inertial
reference frames as well as a symmetry group of coordinate
transformations.

By contrast with $(\ref{SectionVI.2})$-$(\ref{SectionVI.4})$, from
$(\ref{SectionVI.6})$ the anisotropy of the global spacetime ${\cal
R}_{j}(3,3)$ could not be explicitly seen. In fact, according to
$(\ref{SectionVI.6})$ alone the elementary part $P_{jl}$ would be
given in the global spacetime ${\cal R}_{j}(3,3)$ with the signature
$(-, -, -, +, +, +)$ and characterized by three space coordinates
and three time coordinates.

Obviously, it is unusual through the ordinary senses to experience
three dimensions of time. Yet, as far as the description is
concerned, three time coordinates as well as three space coordinates
are determined by three levels of the hierarchical structure.
Moreover, the three dimensions of time and thus the possibility to
travel in time is a result of the fact that in the observation of
the hierarchical structure it is possible not only to consider the
local spacetime of any elementary part, but also change the focus
from one elementary part to another and thus travel in time.

Notably, unlike the hierarchical structure, the representation does
not tell the formation story. It simply does not have information 
about the formation, order and connections between the elementary 
parts. Yet, the representation could be effective as long as through 
the form of the laws the process would be harnessed by understanding 
the elementary parts in terms of the same space and time variables.

To recognize features of familiar global spacetimes let us first
consider a representation of the process in terms of a
four-dimensional global spacetime.

The representation can be obtained as a special case of ${\cal
R}_{j}(3,3), j = 1,...,16$, where the time coordinates of the
elementary parts $P_{j1}, P_{j2}$ and $P_{j3}$ are processed by
using the same time variable. Specifically, in this case a
two-dimensional Euclidean space ${\cal R}_{jl}, j = 1,...,16, \ l =
1,2,3$ with an orthogonal basis of unit vectors ${\bf u}_{jl}$ and
${\bf w}_{j}$ is used to represent the space and time coordinates of
the elementary part $P_{jl}$ by a vector function
\begin{equation}
\label{SectionVI.7} {\bf R}_{jl}(t) = \sqrt{-1}X_{jl}(t){\bf u}_{jl}
+ T_{jl}(t){\bf w}_{j}, \ \ \ t_{j-1} \leq t \leq t_{j}.
\end{equation}

We can see that in comparison with $(\ref{SectionVI.1})$ in
$(\ref{SectionVI.7})$ the unit vector ${\bf w}_{j}$ for the time
coordinate does not depend on the level of the elementary part and
is the same for the Euclidean spaces ${\cal R}_{j1}, {\cal R}_{j2}$
and ${\cal R}_{j3}$.

The direct product
$$
{\cal R}_{j}(3,1) = {\cal R}_{j1} \times {\cal R}_{j2} \times {\cal
R}_{j3}
$$
gives a common stage to the elementary parts $P_{j1}, P_{j2}$ and
$P_{j3}$ and, similar to ${\cal R}_{j}(3,3)$, could be an effective
representation of the process in terms of a four-dimensional global
spacetime with the signature $(-, -, - , +)$, where the elementary
parts $P_{j1}, P_{j2}$ and $P_{j3}$ are characterized by three space
coordinates and one time coordinate.

In particular, in this case the invariant $(\ref{SectionVI.5})$
takes the form
$$
(\frac{dT_{jl}'(t)}{dt})^{2} - (\frac{dX_{jl1}'(t)}{dt})^{2} -
(\frac{dX_{jl2}'(t)}{dt})^{2} - (\frac{dX_{jl3}'(t)}{dt})^{2}
$$
\begin{equation}
\label{SectionVI.8} = (\frac{dT_{jl}(t)}{dt})^{2} -
(\frac{dX_{jl}(t)}{dt})^{2} = 1, \ \ \ t_{j-1} \leq t \leq t_{j}
\end{equation}
and the representation exists as long as it is supported by an
equivalence class of inertial reference frames with the coordinate
transformations preserving the invariant $(\ref{SectionVI.8})$,
where $X_{jl1}'(t), X_{jl2}'(t), X_{jl3}'(t)$ are the space
coordinates and $ T_{jl}'(t)$ is the time coordinate of the
elementary part $P_{jl}$ in such a frame of reference.

The character of the expression $(\ref{SectionVI.8})$ suggests to
consider possible connections of the global spacetime ${\cal
R}_{j}(3,1)$ with general relativity. In particular, while in
general relativity tangent spaces have Minkowskian geometry, in the
representation, by using linear approximations $(\ref{SectionV.16})$
preserving the sum of the energies of the elementary parts, we can
write the invariant $(\ref{SectionVI.8})$ as
$$
\Delta \breve T_{jl}'^{2} - \Delta \breve X_{jl1}'^{2} - \Delta
\breve X_{jl2}'^{2} - \Delta \breve X_{jl3}'^{2}
$$
$$
= \Delta \breve T_{jl}^{2} - \Delta \breve X_{jl}^{2} =
\varepsilon^{2}
$$
and recognize familiar features of the Lorentz transformations in
four-dimensional Minkowski space of special relativity.

Now, let us discuss possible connections of the invariants
$(\ref{SectionVI.5})$ and $(\ref{SectionVI.8})$ with Lie groups.
This might allow, in view of the developments initiated in
\cite{Yang_1}, to consider $(\ref{SectionVI.5})$ in terms of a
Yang-Mills gauge field and its equations, and, since general
relativity is, in fact, the gauge field theory associated with the
symmetry group of Lorentz transformations in Minkowski space
\cite{Utiyama_1}, to derive the Einstein's equations in the case of
$(\ref{SectionVI.8})$.

It is well known that in terms of gauge fields the gravitational
interaction is different from the electromagnetic, strong and weak
interactions \cite{Utiyama_1}, \cite{Yang_2}. In this context it is
interesting to consider whether the description can reveal a
parallel.

In the description the interactions are realized through the prime
integer relations. In particular, it can be interpreted that the
elementary parts are held together in a part by interacting through
the prime integer relation. Furthermore, to see the situation more
traditionally a prime integer relation might be associated with a
gauge field based on the symmetry group of the geometrical pattern.

In particular, let
$$
f_{P_{jl}} = {\cal A}_{jl1}t^{l} + ... + {\cal A}_{jl,l+1} =
\Psi^{[l]}_{1}(t), \ \ \ t_{j-1} \leq t \leq t_{j},
$$
where ${\cal A}_{jl1},..., {\cal A}_{jll}, j = 1,...,16, \ l = 1,2,3$ 
are the quantum numbers of the elementary part $P_{jl}$. To define the 
gauge field between the elementary parts $P_{jl}$ and $P_{j+1,l}$ of 
a part we consider the condition
\begin{equation}
\label{SectionVI.9} f_{P_{jl}} +  f_{{\bf P}_{j,j+1,l}} =
f_{P_{j+1,l}}.
\end{equation}
Since the difference
$$
f_{P_{jl}} - f_{P_{j+1,l}}
$$
$$
= ({\cal A}_{jl1} - {\cal A}_{j+1,l1})t^{l} + ... + ({\cal A}_{jl,l+1}
- {\cal A}_{j+1,l,l+1})
$$
is a polynomial itself
$$
f_{{\bf P}_{j,j+1,l}} = {\cal B}_{j,j+1,l,1}t^{l} + ... + {\cal
B}_{j,j+1,l,l+1},
$$
where
$$
{\cal B}_{j,j+1,li} = {\cal A}_{jli} - {\cal A}_{j+1,li}, \ \ \ i =
1,...,l+1
$$
the gauge field between the elementary parts $P_{jl}$ and
$P_{j+1,l}$ can be associated with an elementary part ${\bf
P}_{j,j+1,l}$, but of a different type.

To be specific, by contrast with the elementary parts $P_{jl}$ and
$P_{j+1,l}$, which experience the gauge field, the elementary part
${\bf P}_{j,j+1,l}$ communicates the field. We may view the field
between the elementary parts $P_{jl}$ and $P_{j+1,l}$ as the
exchange of the elementary part ${\bf P}_{j,j+1,l}$ and say that the
gauge field is required, when the global symmetry of the geometrical
pattern is converted into the local symmetry $(\ref{SectionVI.9})$.

Thus, in the description we can identify two types of elementary
parts with two different roles. First, there are the elementary
parts that experience fields and second, there are the elementary
parts that communicate the fields. Remarkably, as equation
$(\ref{SectionVI.9})$ shows, all elementary parts, in spite of their
differences, are naturally united. It is also important to note that
through three levels of the hierarchical structure we have three
different symmetry groups, which are connected by the
self-organization process.

Importantly, in identifying a gauge field, corresponding to the
gravitational interaction, we might associate it with the symmetry
based on the conservation of the invariant $(\ref{SectionVI.8})$,
rather than the symmetries of the geometrical patterns.

In fact, the invariant $(\ref{SectionVI.8})$, determining the form
the laws of the process take in the global spacetime ${\cal
R}_{j}(3,1)$, appears as a placeholder of the connection between the
space and time coordinates of an elementary part and its character
through the boundary curve. While the invariant is of a general
character, in the representation it has only to work for those
curves that through the correspondence with the prime integer
relations provide the casual links for the elementary parts.

Therefore, as the preservation of the invariant
$(\ref{SectionVI.8})$ might define a gauge field, then through the
invariant's encoding of the connection between the space and time
coordinates we could interpret the gauge field in terms of the
gravitational interaction. Because of the character of the symmetry,
the gravitational interaction would be different from the other
gauge interactions.

Moreover, we can draw some parallels with the Einstein's equations
directly.

First, the connection between the local spacetime and the energy of
the elementary part lies at the core of the description. Indeed, by
the condition $(\ref{SectionV.17})$
\begin{equation}
\label{SectionVI.10} \vert X_{jl}(t) \vert =
{\cal E}_{jl}(t), \ \ \
t_{j-1} \leq t \leq t_{j},
\end{equation}
where $j = 1,...,16, \ l = 1,2,3$ the local spacetime and the energy 
simply represent two different facets of the geometrical pattern and 
thus the prime integer relation itself. The connection can be further 
specified by a metric information about the local spacetime. For example, 
at level $1$
$$
X_{j1}(t) = {\cal A}_{j11}t + {\cal A}_{j12}, \ \ \ t_{j-1} \leq t
\leq t_{j}
$$
and thus we can rewrite the condition $(\ref{SectionVI.10})$ as
\begin{equation}
\label{SectionVI.11}
\vert {\cal A}_{j11}t + {\cal A}_{j12} \vert = {\cal
E}_{j1}(t), \ \ \ t_{j-1} \leq t \leq t_{j},
\end{equation}
where the coefficient ${\cal A}_{j11}$, in view of the coefficient
${\cal A}_{j12}$, provides the metric information about the local
spacetime of the elementary part $P_{j1}$. In fact, the condition
$(\ref{SectionVI.11})$ can be seen as an equation of the geodesic of
the elementary part $P_{j1}$ establishing a precise correspondence
between its spacetime and energy.

Second, the description may interpret the gravitational interaction
similar to the interpretation of the Einstein's equations. In
particular, by using
$$
{\bf R}_{jl}(t) = T_{jl}(t){\bf w}_{jl} + \sqrt{-1}X_{jl}(t){\bf
u}_{jl} =
$$
$$
= \int_{t_{j-1}}^{t}\sqrt{1 + (\frac{dX_{jl}(t')}{dt'})^{2}}dt'{\bf
w}_{jl}
$$
\begin{equation}
\label{SectionVI.12} + \sqrt{-1}X_{jl}(t){\bf u}_{jl}, \ \ \ t_{j-1}
\leq t \leq t_{j},
\end{equation}
in the global spacetime ${\cal R}_{j}(3,3)$ we can define the
velocity of an elementary part $P_{jl}, j = 1,...,16, \ l = 1,2,3$
$$
\frac{d{\bf R}_{jl}(t)}{dt} = \sqrt{1 +
(\frac{dX_{jl}(t)}{dt})^{2}}{\bf w}_{jl} +
\sqrt{-1}\frac{dX_{jl}(t)}{dt}{\bf u}_{jl}
$$
with respect to the reference frame specified by the unit vectors
${\bf w}_{jl}$ and ${\bf u}_{jl}$ and obtain
$$
\vert \frac{d{\bf R}_{jl}(t)}{dt} \vert
$$
\begin{equation}
\label{SectionVI.13} = \sqrt{1 + (\frac{dX_{jl}(t)}{dt})^{2} +
(\sqrt{-1})^{2}(\frac{dX_{jl}(t)}{dt})^{2}} = 1.
\end{equation}

The result $(\ref{SectionVI.13})$ is interesting to be commented.
Namely, the velocity
\begin{equation}
\label{SectionVI.14} \vert \frac{d{\bf R}_{jl}(t)}{dt} \vert = \vert
\widetilde{V}_{jl}(t) \vert = 1
\end{equation}
is a dimensionless quantity and, by defining the dimensional
velocity $\widetilde{v}_{jl}(t)$ of the elementary part $P_{jl}$
with the view on $(\ref{SectionV.5})$, the condition
$(\ref{SectionVI.14})$ can be written as
$$
\vert \frac{d{\bf R}_{jl}(t)}{dt} \vert = \vert
\widetilde{V}_{jl}(t) \vert = \vert \frac{\widetilde{v}_{jl}(t)}{c}
\vert = 1
$$
and thus
\begin{equation}
\label{SectionVI.15} \vert \widetilde{v}_{jl} \vert = \vert
\widetilde{v}_{jl}(t) \vert = c,
\end{equation}
which means that the elementary part $P_{jl}$ moves with the speed
of light $c$.

Now, we can recall from $(\ref{SectionV.8})$ that in the
hierarchical network the velocity $v_{j1}$ of the elementary part
$P_{j1}$ satisfies the condition
\begin{equation}
\label{SectionVI.16} \vert \frac{v_{j1}}{c} \vert =
\frac{\chi_{0}}{\sqrt{\chi_{min}^{2} + \chi^{2}_{0}}} < 1
\end{equation}
and hence $\vert v_{j1} \vert < c$. Notably, the condition
$(\ref{SectionVI.16})$ gives $\vert v_{j1} \vert = c$ when
$\chi_{min} = 0$ and shows that $\vert v_{j1} \vert$ is very close
to $c$ when $\chi_{min} << \chi_{0}$.

Therefore, since the space and time coordinates of the elementary
part $P_{jl}$ are given by $(\ref{SectionVI.12})$ without
information about the process and the connection between the levels
in particular, we may say that in the global spacetime ${\cal
R}_{j}(3,3)$ the true character of $(\ref{SectionVI.16})$ become
hidden in $(\ref{SectionVI.15})$.

Next, we obtain the acceleration of the elementary part $P_{jl}$
$$
\frac{d^{2}{\bf R}_{jl}(t)}{dt^{2}} =
\frac{\frac{dX_{jl}(t)}{dt}\frac{d^{2}X_{jl}(t)}{dt^{2}}} {\sqrt{1 +
(\frac{dX_{jl}(t)}{dt})^{2}}}{\bf w}_{jl} +
\sqrt{-1}\frac{d^{2}X_{jl}(t)}{dt^{2}}{\bf u}_{jl}
$$
as well as its tangential and normal components in the global
spacetime ${\cal R}_{j}(3,3)$.

By using the dot product and $(\ref{SectionVI.13})$, for the
tangential component $\eta_{jl}$ we get
$$
\eta_{jl}(t) = \frac{\frac{d{\bf R}_{jl}(t)}{dt} {\bf \cdot}
\frac{d^{2}{\bf R}_{jl}(t)}{dt^{2}}}{\vert \frac{d{\bf
R}_{jl}(t)}{dt} \vert}
$$
$$
= \sqrt{1 + (\frac{dX_{jl}(t)}{dt})^{2}} \cdot
\frac{\frac{dX_{jl}(t)}{dt}\frac{d^{2}X_{jl}(t)}{dt^{2}}}{\sqrt{1 +
(\frac{dX_{jl}(t)}{dt})^{2}}}
$$
$$
+ \sqrt{-1}\frac{dX_{jl}(t)}{dt} \cdot
\sqrt{-1}\frac{d^{2}X_{jl}(t)}{dt^{2}}
$$
\begin{equation}
\label{SectionVI.17} =
\frac{dX_{jl}(t)}{dt}\frac{d^{2}X_{jl}(t)}{dt^{2}} -
\frac{dX_{jl}(t)}{dt}\frac{d^{2}X_{jl}(t)}{dt^{2}} = 0.
\end{equation}

In its turn, by using the cross product and $(\ref{SectionVI.13})$,
for the normal component $\mu_{jl}$ we have
\begin{equation}
\label{SectionVI.18} \mu_{jl}(t) = \kappa_{jl}(t) \vert \frac{d{\bf
R}_{jl}(t)}{dt} \vert^{2} = \kappa_{jl}(t),
\end{equation}
where
$$
\kappa_{jl}(t) = \frac{\vert \frac{d{\bf R}_{jl}(t)}{dt} \times
\frac{d^{2}{\bf R}_{jl}(t)}{dt^{2}} \vert }{\vert \frac{d{\bf
R}_{jl}(t)}{dt} \vert^{3}}
$$
is the curvature of the spacetime. Since
$$
\frac{d{\bf R}_{jl}(t)}{dt} \times \frac{d^{2}{\bf
R}_{jl}(t)}{dt^{2}}
$$
$$
= ({\sqrt{1 + (\frac{dX_{jl}(t)}{dt})^{2}}} \cdot
\sqrt{-1}\frac{d^{2}X_{jl}(t)}{dt^{2}}
$$
$$
-\sqrt{-1}\frac{dX_{jl}(t)}{dt} \cdot
\frac{\frac{dX_{jl}(t)}{dt}\frac{d^{2}X_{jl}(t)}{dt^{2}}}{\sqrt{1 +
(\frac{dX_{jl}(t)}{dt})^{2}}}){\bf k}_{jl}
$$
$$
= (\sqrt{-1}\frac{\frac{d^{2}X_{jl}(t)}{dt^{2}}}{\sqrt{1 +
(\frac{dX_{jl}(t)}{dt})^{2}}}){\bf k}_{jl}
$$
for the curvature of the spacetime we obtain
$$
\kappa_{jl}(t) =
\frac{\sqrt{-1}\frac{d^{2}X_{jl}(t)}{dt^{2}}}{\sqrt{1 +
(\frac{dX_{jl}(t)}{dt})^{2}}},
$$
where ${\bf k}_{jl} = {\bf w}_{jl} \times {\bf u}_{jl}$.

In summary, since, according to $(\ref{SectionVI.17})$, the
tangential component $\eta_{jl} = 0$, the acceleration of the
elementary part $P_{jl}$ through the normal component, as
$(\ref{SectionVI.18})$ shows, is fully determined by the curvature
of the spacetime
$$
\mu_{jl}(t) = \kappa_{jl}(t).
$$

Following the Einstein's principle of equivalence that gravity
equals acceleration \cite{Einstein_2}, we could associate the
acceleration with the gravitational interaction and find that the
gravitation would be equivalent to the curvature of the spacetime.

Now, we consider a representation of the process in terms of a 
three-dimensional global spacetime, where space and time are, in fact, 
independent of each other. The main difference in this case is that in the
representation of the boundary curve the time coordinate of the
elementary part become associated with the parameter $t$. As a
result, we can obtain an Euclidean three dimensional space with the
time running independently and in the same manner for all elementary
parts.

In particular, in the representation the space coordinate of an
elementary part $P_{jl}, j = 1,...,16, \ l = 1,2,3$ is specified by
a vector function
\begin{equation}
\label{SectionVI.19} {\bf R}_{jl}(t) = X_{jl}(t_{j-1} + t){\bf
u}_{l}, \ \ \ t \in [0, \varepsilon]
\end{equation}
in a one-dimensional Euclidean space ${\cal R}_{l}, l = 1,2,3$,
where ${\bf u}_{l}$ is the unit vector, while the time coordinate is
given by
\begin{equation}
\label{SectionVI.20} T_{jl}(t_{j-1} + t)= t, \ \ \ t \in [0,
\varepsilon].
\end{equation}

The direct product
$$
{\cal R}(3,0) = {\cal R}_{1} \times {\cal R}_{2} \times {\cal R}_{3}
$$
provides a common stage to the elementary parts, where an elementary
part can be characterized by three space coordinates and time.

As the boundary curve of the elementary part $P_{jl}$ become represented
by the conditions $(\ref{SectionVI.19})$ and $(\ref{SectionVI.20})$,
the quantum of the laws carried by the elementary part takes the form
of a curve in a two-dimensional Euclidean plane. This form will be
preserved in the global spacetime ${\cal R}(3,0)$, as long as it is
supported by an equivalence class of inertial reference frames with
the coordinate transformations leaving the expression invariant
\begin{equation}
\label{SectionVI.21} dX_{jl}^{2}(t) = dX_{jl1}'^{2}(t) +
dX_{jl2}'^{2}(t) + dX_{jl3}'^{2}(t),
\end{equation}
where $X_{jl1}'(t), X_{jl2}'(t), X_{jl3}'(t), \ t \in [t_{j-1},
t_{j}]$ are the space coordinates of the elementary part $P_{jl}$ in
such a frame of reference. Through the character of the invariant
$(\ref{SectionVI.21})$ familiar features of the Galilean
transformations in three dimensional Euclidean space of Newtonian
mechanics can be recognized.

Now let us consider how the loss of information about the process
might determine the understanding of the elementary parts in an
effective representation.

Since in a global spacetime the elementary parts could be perceived
by the trajectories in the first place, they would initially become
the main subject of the understanding. This might be resolved by
finding equations of motion with the parameters adjusted by
experiments precisely for the equations to work.

With the equations of motion producing the trajectories in agreement
with observation, the power of equation would be recognized to
establish one master equation to unify them all. However, the
equations would resist, because the elementary parts are, in fact,
unified by the self-organization process rather than any single
equation.

Since the trajectory of an elementary part is encoded by the
geometrical pattern, some of the parameters would be specific to the
geometrical pattern, while the others could be more universal to
reflect its belonging to the hierarchical structure of geometrical
patterns. As the universal parameters would be relevant to all
elementary parts, they might be especially distinguished and called
"constants of nature". Once the parameters and "constants of nature"
could be calculated, it would be then important to understand where
they all come from and why they have the values as they do.

Besides, it would be like a mystery to find out that some of the
parameters are actually fine-tuned, i.e., each digit in the value
must stand as it is and not be even slightly otherwise. Moreover,
although digits in the value of a fine-tuned parameter might look
randomly placed, yet surprisingly each digit is strictly determined.
In fact, if the value of the parameter were varied just a bit, then
systems made of the elementary parts would cease to exist.

Moreover, as the elementary parts are characterized by the energies
and quantum numbers, which are preserved under certain conditions,
the corresponding laws of conservation would become one of the main
pillars in the understanding of the elementary parts.

Furthermore, since an elementary part at level $l = 1,2,3$ is
specified by $l$ quantum numbers, three generations of the
elementary parts could be found. Due to the symmetries of the
geometrical patterns of the parts at a level, the elementary parts
of a generation would be characterized by a symmetry group. And,
because the geometrical patterns at the levels are all connected by
the process, it could be revealed that the symmetry groups of three
generations of the elementary parts are, in fact, connected thus
tempting to unify the symmetries by one large symmetry.

With the progress made so far it would be possible to contemplate
why there exist three space dimensions and three generations of the
elementary parts and whether these facts might be connected. The
role of time would be especially puzzling. For example, why there
are three space dimensions and only one dimension of time and
whether other combinations could be also possible.

Notably, the understanding of the elementary parts in an effective
representation resonates with inquiries on a number of fundamental
issues such as the unification of forces, constants of nature, the
standard model of elementary particles and the nature of space and
time itself \cite{Einstein_4}-\cite{Davies_1}.

\section{Possible Implications}

In the previous sections we have presented results based on the
description of complex systems in terms of self-organization
processes of prime integer relations. Although only one
self-organization process has been considered, yet, we have obtained
a first resolution picture of the hierarchical network revealing
remarkable features of the description.

In particular, the description not only combines key features of
quantum mechanics and general relativity to appear as a potential
candidate for their unification
\cite{Korotkikh_7}-\cite{Korotkikh_9}, but also presents something
that might constitute a new physics. Namely, it raises the
possibility that the law of conservation of energy and the second
law of thermodynamics can loose their generality and become
different manifestations of a more fundamental entity, i.e, the
self-organization processes of prime integer relations and thus
arithmetic.

Moreover, the elementary parts of the correlation structure act as
the carriers of the laws of arithmetic with each single elementary
part carrying its own quantum of the laws. This opens an important 
perspective to consider elementary parts in the hierarchical network 
as quanta to construct different laws and thus proposes the hierarchical
network as a source of laws. In particular, like the transformation
of energy into different forms, the description suggests that the
laws of arithmetic of the hierarchical network could be transformed
into different forms by constructing global spacetimes.

Furthermore, the description demonstrates features of quantum
entanglement \cite{Einstein_5}-\cite{Gisin_1}, backward causality
\cite{Wheeler_2}-\cite{Wickes_1} and possible extensions and
interpretations of physical theories
\cite{Sakharov_1}-\cite{Davies_2}. In addition, the character of
nonlocality and reality it advocates finds parallels in
philosophical, religious and mystical teachings
\cite{Capra_1}-\cite{Radin_1} as well as psychic phenomena
\cite{Radin_1},\cite{Radin_2}.

At the same time, there is one feature that crucially distinguishes
the description. Based on the integers and controlled by arithmetic
only
\medskip

{\it the description has an utterly unique potential to complete the
quest for the fundamental laws of nature}.
\medskip

In view of this unique potential we discuss possible implications 
of the results as they may provide the answers to many key questions.

{\it First}, the question about the possible ultimate building
blocks of nature has been one of the greatest questions of all time.
Ever since Pythagoras integers have been believed to be a likely
candidate for this role. The description may fulfil the expectation.
Namely, in our description the integers appear as the ultimate building
blocks of the self-organization processes in the construction of the
hierarchical network of prime integer relations. The description
suggests the hierarchical network as a new arena for understanding
and dealing with complex systems.

Remarkably, the description comes up with an answer to the question
about the elementary particles. From its perspective the elementary
parts or particles are all encoded and interconnected by the
self-organization processes of prime integer relations. In
particular, an elementary particle, as a part of a correlation
structure, is entirely characterized by a two-dimensional
geometrical pattern, which, in its turn, is specified by the
boundary curve.

Therefore, in the description an elementary particle can be seen as
a curve given by a polynomial with all its coefficients as the
quantum numbers of the elementary particle, except the last one.
Notably, the quantum numbers of the elementary particles of a
correlation structure are all conserved.

As a result, the description gives the clear message that all
elementary particles may be already represented in the hierarchical
network with the structures and parameters completely determined by
arithmetic through the processes. In other words, no matter how
powerful colliders can be, no elementary particles could be found,
unless they would be encoded through the hierarchical network.

Furthermore, in understanding the mechanism the elementary particles
may acquire their masses, be aware that the mass of an elementary
particle may be fully determined by the area under the boundary
curve, which is absolutely fixed by arithmetic and thus can not be
changed at all.

{\it Second}, the description suggests that the forces of nature
could be unified. Namely, in the realm of the hierarchical network
all forces are managed by the single "force" - arithmetic to serve
the special purpose: to hold the parts of a system together and
possibly drive its formation to make the system more complex.
Therefore, in the description the forces do not exist separately,
but through the self-organization processes of prime integer
relations are all unified and controlled to work coherently in the
preservation and formation of complex systems.

Notably, in the hierarchical network the information about a complex
system is fully encoded by the position, which determines the forces
acting on the system, its physical constants and parameters.

{\it Third}, the description raises the possibility of a deeper
reality with space and time as its effective representations.
Importantly, the description makes the reality comprehensible by
providing its mathematical structure, i.e., the hierarchical network
of prime integer relations. This would allow to develop theoretical
and practical tools to live and operate in this new reality. Because
the hierarchical network is based on integers only and thus
irreducible, the search for a more deeper reality might even become
irrelevant.

Where would we be in that possible reality? It seems likely that, as
a starting point of reference, the standard model of elementary
particles might be useful. Indeed, when self-organization processes
take place in the hierarchical network they could encode certain
elementary particles. Therefore, representing the standard model in
terms of the description would help to identify the underlying
processes and thus the position in the hierarchical network. For
this purpose one large symmetry of the hierarchical network may be
used to accommodate the elementary particles through their
symmetries. Yet to navigate to the position it seems that particle
accelerators would be needed as compasses in the new reality.

Since things exist in the hierarchical network as integrated parts
of complex systems, the position could reveal a larger system and a
bigger process.

Some properties of the new reality are especially appealing. For
example, in space and time the distances between systems are
important and can be frustratingly large to establish the
connection. Moreover, so far it remains unknown whether past,
present and future might be connected all at once. However,
according to the description systems can be instantaneously
connected and united irrespective of how far they may be apart in
space and time. Moreover, in the hierarchical network a complex
system could be managed through self-organization processes as a
whole with its possible pasts, presents and futures all at once.

The description promises two new sources of energy and laws. In
particular, in the description arithmetic controls energy and thus
the energy must be conserved or generated exactly in the amount
determined by arithmetic. Therefore, the description suggests a new
source of energy controlled through the processes. This source may
be already observed by dark energy and matter
\cite{Riess_1},\cite{Perlmutter_1}, yet, it would be a completely
different story to be able to use it with all technological
consequences.

Furthermore, the hierarchical network could be used as a source of
laws to achieve different objectives. In particular, for a given
objective the hierarchical network could be used to generate
self-organization processes providing relevant laws of arithmetic to
be processed into the required form by constructing a corresponding
global spacetime. Remarkably, in managing this source space and time
would be manipulated as dynamic variables of the new reality.

{\it Fourth}, paradoxically in the world experiencing rapid progress
in science and technology the understanding of the growing
challenges is limited.

For example, the recent economic crisis has sharply revealed that
the financial system shaped and entangled by the globalization into
a complex entity lacks the understanding to clearly see the way out.
Furthermore, the climate change debates going for a rather long time
still have not resulted in a comprehensive agreement. Basically,
there is no understanding of this complex problem, where
experimental evidence could result in arguments convincing to
different parties.

And now it has come to recognize whether the challenges can be the
first signs of a profound transformation, where the current
scientific view of reality operating through space and time may be
no longer good enough to guide us any further. In searching for
alternative the description suggests a deeper reality, where
efficient management of complex systems could be possible
\cite{Korotkikh_10}-\cite{Korotkikh_19}.

Importantly, based on integers only, the description could provide a
common ground to be fully trusted by different parties. Similar,
like nowadays arithmetic is used to count things, we hope, it would
be possible to use arithmetic to deal with things, no matter how
complex they might be.

{\it Fifth}, ever since Socrates and Plato it has been considered
that through the ordinary senses we might only observe projections
of some high-dimensional reality. In this regard, it is remarkable
that the hierarchical network can be seen as a high-dimensional
reality, while global spacetimes appear as its effective
projections.

Moreover, in the processing of the hierarchical network the
resulting effective representation could be defined by the
advantages it may give to the observing system. In the context of
the mind-matter problem this allows us to speculate that in the
processing of the hierarchical network for the time being the mind
might be divinely determined to sense three dimensions of space and
one dimension of time.

Furthermore, a system could be programmed to different global
spacetimes and achieve desired objectives without realizing the code
itself, i.e, the hierarchical network as a deeper reality. We
believe that in providing the code the description opens a way to
make a transformation for a new role humans have to play.

{\it Finally} and perhaps most remarkably, although the integers are
truly fundamental, yet, they are simply products of human thinking.
\medskip
\medskip

{\it "It is through thought we must raise ourselves, and not through
space and time, which we can never fill. So let us strive to think
well: this is the mainspring of morality."}

\begin{flushright}
--- Blaise Pascal, {\it Pensees}
\end{flushright}
\medskip
\medskip

{\it Acknowledgements.} I am especially thankful to my dearest
parents for the gift of life, my beautiful wife Galina for
enlightening my soul and children Grigori and Maria for their
patience and love.

\bibliography{apssamp}

\end{document}